\documentclass[journal]{IEEEtran}

\ifCLASSINFOpdf

\else

\fi

\usepackage{float}
\usepackage{cite}
\usepackage{amsmath,amssymb,amsfonts}
\usepackage{algorithmic}
\usepackage{graphicx}
\usepackage{textcomp}
\usepackage{xcolor}
\usepackage{gensymb}
\usepackage{makecell}
\usepackage{stfloats}
\usepackage{amsthm}
\usepackage{amssymb}
\usepackage{amsmath}
\usepackage{mathtools}
\usepackage{extarrows}
\usepackage{subfigure}
\usepackage{tablefootnote}
\usepackage{threeparttable}
\usepackage{footnote}
\usepackage{color, soul, framed}

\usepackage{subfigure}

\begin{document}

\title{FDA-MIMO-based Integrated Sensing and Communication System with Frequency Offset Permutation Index Modulation}

\author{Jiangwei~Jian,~\IEEEmembership{Student Member,~IEEE,}
	Qimao~Huang,
	Bang~Huang,~\IEEEmembership{Graduate Student Member,~IEEE}
	and~Wen-Qin~Wang,~\IEEEmembership{Senior Member,~IEEE}
	\thanks{This work was supported by National Natural Science Foundation of China 62171092. (Corresponding author: Wen-Qin Wang)}
	\thanks{Jiangwei Jian, Wen-Qin Wang, and Bang Huang are with School of Information and Communication Engineering, University of Electronic Science and Technology of China, Chengdu, 611731, P. R. China. (Email: jianjiangwei@std.uestc.edu.cn; huangbang@std.uestc.edu.cn; wqwang@uestc.edu.cn).
		
	Qimao Huang is with School of Physics, University of Electronic	Science and Technology of China, Chengdu, 611731, P. R. China. (Email: huangqimao@std.uestc.edu.cn). }	
}


\maketitle

\begin{abstract}	
Considering that frequency diverse array multiple-input multiple-output (FDA-MIMO) possesses extra range information to enhance sensing performance, this paper explores the FDA-MIMO-based integrated sensing and communication (ISAC) system. To reinforce the system communication capability, we propose the frequency offset permutation index modulation (FOPIM) scheme, which conveys extra information bits by selecting and permutating frequency offsets from a frequency offsets pool. For the system communication sub-functionality, considering the fact that the traditional maximum likelihood detection method suffers from high complexity and bit error rate (BER), the maximum likelihood-based two-stage detection (MLTSD) approach is presented to overcome this issue. For the system sensing sub-function, we employ the two-step maximum likelihood estimator (TSMLE) to stepwise estimate the angle and range of the interested target. Furthermore, we derive the closed-form expressions for the tight upper bound on the communication BER, along with the sensing Cramér-Rao bound (CRB). The simulation results validate the theoretical analysis, demonstrating that the proposed system exhibits lower BER and superior range resolution than independent MIMO communication and MIMO sensing modules.

\end{abstract}


\IEEEpeerreviewmaketitle

\section{Introduction}
\IEEEPARstart{I}{ntegrated} sensing and communication (ISAC) technique has attracted tremendous recent attention, for its capability to perform communication and sensing functions in the same platform \cite{ma2020joint,liu2022integrated,chen2022generalized}. ISAC technique can operate at different integration levels, from non-overlapping resource allocation \cite{d2019communications,zhang2022time,ye2022low} for communication and sensing to fully unified waveforms \cite{liu2018mu,liu2020joint}. Unlike resource allocation ISAC, which wastes resources, the fully unified waveform ISAC reduces cost and complexity by utilizing all resources for communication and sensing simultaneously \cite{liu2022integrated}. Existing fully unified waveform ISAC research can be categorized into three main areas: Communication-centric design (CCD), sensing-centric design (SCD), and joint ISAC design \cite{liu2022integrated,liu2018mu,ma2021spatial,sturm2011waveform}.

The CCD-based ISAC methods aim to enable additional sensing capability for existing communication systems, relying primarily on orthogonal frequency division multiplexing (OFDM) waveforms \cite{liu2022integrated,sturm2011waveform,johnston2022mimo,temiz2021dual}. The OFDM waveform was first designed for ISAC by Sturm \textit{et. al} in 2009 \cite{sturm2011waveform}. Their approach incorporated the fast Fourier transform (FFT) and Inverse FFT (IFFT) algorithms to estimate velocity and distance, achieving favorable sensing and communication performance. Subsequently, MIMO was combined with OFDM in \cite{johnston2022mimo}, whose waveform was optimized under communication performance constraints to guarantee a high communication quality of service. On the other hand, an MIMO OFDM-based ISAC technique for the uplink was proposed in \cite{temiz2021dual}, which enabled users to sense short-range targets during uplink communication simultaneously. However, \cite{johnston2022mimo,temiz2021dual} suffers from intercarrier interference (ICI). To address this, \cite{keskin2021mimo} utilized the estimated carrier frequency offset to convert ICI from foe to friend, thereby improving the system sensing performance. 

In contrast to CCD, SCD-based ISAC schemes consider the target sensing as its primary function \cite{liu2022integrated,ma2020joint}. Early sensing-centric designs focused on chirp waveforms \cite{roberton2003integrated,saddik2007ultra}, offering an excellent range estimate performance through pulse compression. To achieve higher communication rates, a promising approach is to embed communication information through index modulation (IM) techniques \cite{ma2021spatial}. In \cite{hassanien2015dual}, the communication bits were embedded by controlling the amplitude of the beampattern sidelobe. Later, \cite{wang2018dual} leveraged MIMO's spatial diversity indexing to convey extra information bits. However, the sparse array used by \cite{wang2018dual} resulted in degraded sensing performance. To address this, \cite{huang2020majorcom} proposed a carrier-agile phased array-based ISAC scheme, which maps information bits into waveforms and frequencies to obtain improved sensing performance. Further, \cite{ma2021frac} employed the joint index modulation of frequencies, antennas, and waveforms, obtaining lower BER and enhanced target estimation accuracy.

However, the CCD and SCD suffer from weak sensing and communication capability, respectively \cite{liu2022integrated}. To balance two conflicts, the joint design-based ISAC method has attracted attention \cite{chen2021joint,keskin2021limited,liu2018toward,liu2021cramer,liu2022transmit}. Specifically, in the single-user scenario, \cite{keskin2021limited} attained a favorable performance trade-off between sensing and communications by optimizing subcarrier power and introducing a feedforward channel. For multi-user scenarios, \cite{liu2018toward} designed a robust precoding strategy to maximize detection probability while maintaining communication quality. Ref. \cite{liu2021cramer} further extended the work to achieve optimal sensing performance while meeting communication requirements using the Cramér-Rao bound (CRB) constraint. Moreover, ref. \cite{liu2022transmit} explored covariance optimization criteria for waveform design and interference suppression.

Nevertheless, the aforementioned ISAC systems primarily rely on phased arrays (PAs), which can only generate angle-dependent beampatterns but lack distance-dependent capabilities. This limitation restricts their sensing performance. The frequency diverse array (FDA) technique can surmount this constraint by employing a frequency offset across the neighboring array elements to generate the angle- and range-dependent beampattern \cite{antonik2006frequency,wang2016overview,lan2020glrt}. However, the FDA exhibits inherent time variation issues \cite{wang2015frequency}. To address this, FDA was combined with MIMO by \cite{sammartino2013frequency}, named FDA-MIMO, which has shown superior performance compared to PA in radar applications. Specifically, refs. \cite{xu2015joint,xiong2017fda} utilized FDA-MIMO for target angle and range estimation, demonstrating FDA's superior range super-resolution capability compared to PA-based radars. Additionally, the benefits of FDA-MIMO radar in target detection, mainbeam deceptive interference suppression, and range clutter suppression are detailed in \cite{huang2021adaptive,lan2020suppression,xu2015range}.

Moreover, the FDA can also benefit communication. Its earliest application in communication was the physical layer security, facilitating secure communication in both angle and range dimensions \cite{ji2019physical,jian2023physical}. Subsequently, several works have incorporated IM techniques into FDA to improve communication rates \cite{basit2021fda,nusenu2020space,qiu2020multi,jian2023mimo}. Notably, the FDA-based frequency offsets index modulation (FOIM) scheme was proposed in \cite{jian2023mimo}, which selected frequency offsets from a pool to carry extra bits. However, the FOIM method failed to utilize the frequency permutation and thus limited the communication rate. This limitation motivates us to further employ the FDA frequency permutation to enchance the communication performance in this paper. 

Due to the FDA's notable advantages in radar and communication, integrating its communication and sensing functions into a unified platform has captured the interest of certain researchers \cite{ji2018dual,nusenu2018time,basit2019range,li2023joint}. Nusenu \textit{et. al} introduced an FDA-MIMO radar-centric ISAC system \cite{nusenu2018time}, which implemented communication by embedding information bits through the spread sequence. 
To minimize the impact of the communication function on sensing performance, \cite{basit2019range} conveyed communication information by varying null depths without altering the width of the mainlobe. To achieve higher communication rates, \cite{basit2019range,li2023joint} were reported for embedding extra index bits in the beampattern and waveform domains of the FDA, respectively.
However, these designs overlooked transmit beamforming optimization and thus deducing the sensing performance. To address this problem, \cite{zhou2021performance} suggested inserting weighted phase-modulated communication signals into FDA-MIMO radar waveforms, achieving an optimal trade-off between radar and communication performance. 

However, existing FDA-MIMO-based ISAC schemes underutilize FDA resources in frequency and spatial dimensions for communication performance enhancement. Regarding sensing, most previous work only examined the CRB for angle and range estimation without proposing a dedicated estimation algorithm. Motivated by these limitations, this paper proposes the frequency offset permutation index modulation (FOPIM) scheme for the FDA-MIMO-based ISAC system. The main contributions of our work are listed as follows:
\begin{itemize}

\item [1)]
In this work, we further exploit the communication resources of the FDA in frequency and spatial dimensions and propose the FOPIM scheme to improve the communication rate of the considered ISAC system. The FOPIM scheme conveys additional information bits by selecting and permutating frequency offsets from a frequency offset pool. Moreover, each transmit antenna is assigned a separate quadrature amplitude modulation (QAM) symbol to increase the data rate.

\item [2)]
For the system communication sub-function, the traditional maximum likelihood (ML) decoder face complexity and BER challenges. To tackle this issue, an ML-based two-stage decoding (MLTSD) approach is proposed by utilizing the parallel architecture of the communication receiver. In the first stage, the selected frequency offsets are estimated by searching maximum values of receiver outputs. Then, in the second stage, the antenna index and attached QAM symbol of each frequency offset are estimated by the ML criterion. Moreover, we derive closed-form expressions for the tight upper bound on the system BER.

\item [3)]
For the system sensing sub-function, a suitable receiver structure is designed to mitigate the effect of QAM symbols. Then, the two-step ML estimator (TSMLE) is employed to estimate the target angle and range stepwise. Specifically, we use the ML estimator to estimate the target angle and then utilize this estimated angle as prior information for further target range estimation. Additionally, the closed-form expressions for the system CRB are derived in this paper.

\end{itemize}

The rest of this paper is organized as follows. Section \ref{sc2} proposes the FOPIM for the FDA-MIMO-based ISAC system. Section \ref{sc3} discusses the signal processing methods for the system communication and sensing sub-functions.  Section \ref{sc4} analyzes the theoretical performance of the proposed system, i.e., BER and CRB. Finally, simulation results are provided in section \ref{sc5}, and the paper is concluded in section \ref{sc6}.

$Notations$: $\mathbf{I}_N$ denotes the identity matrix of order $N$. $\lfloor \cdot \rfloor$, $!$ and $\Gamma (\cdot)$ denote the floor function, factorial and Gamma function, respectively. $\mathrm{Z}$ denotes the set of positive integers. $^{T}$ and $^{\dagger}$ stand for the transpose and conjugate operations, respectively. $\mathcal{C} \mathcal{N} (\mu ,\sigma) $ denotes the complex Gaussian distribution with mean $\mu $ and variance $\sigma$. $\left\| \cdot \right\|$ and $\left| \cdot \right|$ represent the Frobenius norm operation and the absolute value operation, respectively. $\odot $ represents the Hadamar product, $C_{k}^{r}=\frac{k!}{r!(k-r)!}$ denotes the binomial coefficient. $\mathrm{Re}\left\{ \cdot \right\} $ and $\mathrm{Im}\left\{ \cdot \right\} $ denote the real part operator and the imaginary part operator, respectively. $\mathrm{diag}\left\{ \cdot \right\} $ denots the diagonal matrix, $*$ denotes the convolution operator.

\section{System Model}
\label{sc2}
We consider the scenario where a co-located FDA-MIMO-based ISAC base station (BS) senses a single target while serving a communication user with $N$ transmit antennas and $M$ receive antennas. The downlink communication user equips $L$ antennas with the perfect channel state information and synchronizes with the BS \cite{basit2021fda}.

\begin{figure}[htp]
	\centering
	\subfigure	{\includegraphics[width=0.4\textwidth]{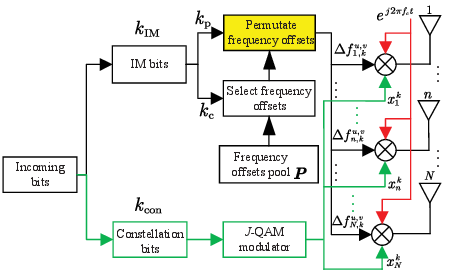}}
	\caption{Block diagram of the proposed FOPIM scheme.}
	\label{FOPIM_diag}
\end{figure}

At the transmitter, the recent FOIM method \cite{jian2023mimo} can be directly deployed in the transmitter to enhance the communication perfromance. However, the FOIM scheme fails to utilize the frequency offsets permutation, thus limiting its communication rate. To overcome this shortcoming, as shown in Fig. \ref{FOPIM_diag}, the FOPIM method is proposed in this paper. Compare to the FOIM approach, the proposed FOPIM scheme adds a frequency offsets permutation module (marked in yellow in Fig. \ref{FOPIM_diag}), achieving a higher communication rate.

Specifically, the proposed FOPIM method divides incoming bits into two essential parts, namely index modulation bits $k_{\mathrm{IM}}$, and constellation bits $k_{\mathrm{con}}$, where $k_{\mathrm{IM}}$ is further divided as frequency offset combination bit $k_{\mathrm{c}}$ and frequency offset permutation bits $k_{\mathrm{p}}$. We set a frequency offset pool $\boldsymbol{P}=[0,\Delta f,\cdots ,\left( P-1 \right) \Delta f],P>N$, and the $N$ transmit frequency offsets are selcted from $\boldsymbol{P}$. Note that $\Delta f$ denotes the frequency increment. One can observe that there exists $C_{P}^{N}$ frequency offsets combinations that carry $k_{\mathrm{c}}=\lfloor \log _2 C_{P}^{N} \rfloor$ bits. Then, the selected $N$ frequency offsets are permutated. In this way, the frequency offset permutation bits are restricted as $k_{\mathrm{p}}=\lfloor \log _2 N! \rfloor$. On the other hand, the constellation bits are modulated into $N$ unit power $J$-ary QAM symbols, where the $n$th constellation symbol is carried by the antenna with the $n$th large frequency offset. 
Therefore, with the same parameter configuration, the FOPIM approach carries $N\log _2J+\lfloor \log _2C_{P}^{N} \rfloor +\lfloor \log _2N! \rfloor$ bits, while the FOIM method can only carry $N\log _2J+\lfloor \log _2C_{P}^{N} \rfloor$ bits \cite{jian2023mimo}. That is, the proposed FOPIM method boosts the spectral efficiency of the FOIM scheme.

In the proposed FOPIM scheme, there are $U=2^{k_{\mathrm{c}}}$ frequency offset combination events and each contains $V=2^{k_{\mathrm{p}}}$ permutation events. Hence, there are $UV$ frequency offset combinatorial permutation events.
Let $\mathbf{f}_{k}^{u,v}=[ \Delta f_{1,k}^{u,v},\cdots ,\Delta f_{n,k}^{u,v},\cdots ,\Delta f_{N,k}^{u,v} ] ^T$ represent the permutated frequency offset vector in the $k$th pulse repetition interval (PRI). Then, the transmit signal of the $n$th antenna at the $k$th PRI is written as 
\begin{equation}\label{s_kn}
	\begin{aligned}
		s_{k}^{n}\left( t \right) =\phi \left( t-\left( k-1 \right) T \right) \sqrt{\frac{P_S}{N}}e^{j2\pi \left( f_c+\Delta f_{n,k}^{u,v} \right) t}x_{n}^{k}	,
	\end{aligned}
\end{equation}
where $x_{n}^{k}$ represents the QAM symbol transmitted by the $n$th antenna in the $k$th PRI. $P_S$ and $f_c$ denote the total transmitter power and the carrier frequency, respectively. $T_p$ and $T$ stand for the pulse width and PRI, respectively. $\phi (t) $ denotes the complex baseband pulse signal with unit energy, i.e., $\int_{T_P}{\phi \left( t \right) \phi ^{\dagger}\left( t \right) dt}=1$, and satisfies the following orthogonal relation \cite{lan2020suppression,ji2019physical,zhou2021performance}
\begin{equation}\label{orthogonality}
	\begin{aligned}
		\int_{T_P}{\phi (t) \phi ^{\dagger}(t) e^{j2\pi \kappa \Delta ft}dt}=\left\{ \begin{array}{r}
			1,\kappa =0\\
			0,\kappa \ne 0\\
		\end{array} \right.  	.
	\end{aligned}
\end{equation}
Note that the orthogonality in \eqref{orthogonality} will be guaranteed when $\Delta f$ satisfies the $\Delta f=\frac{i}{T}, i\in \mathrm{Z}$ condition \cite{basit2021fda,lan2020glrt,senanayake2022frequency}.

\section{System communication and sensing functions}
\label{sc3}
This section discusses the received signal models and signal processing methods for the system's communication and sensing sub-functions.

\subsection{System communication sub-function}

We consider a motionless communication user located at a far-field point $\left( R_C,\theta _C \right)$. Since the transmitted signals possess frequency diversity, the frequency-selective Rayleigh block fading MIMO channel is considered in this paper \cite{bacsar2013orthogonal,jian2023mimo}. 
We stack the channel coefficient matrices corresponding to each frequency offset into a total matrix as $\mathbf{H}=\left[ \mathbf{H}_1,\cdots ,\mathbf{H}_p,\cdots \mathbf{H}_P \right] \in \mathbb{C} ^{L\times NP}$. Note that $\mathbf{H}_p=\left[ \mathbf{h}_{p}^{1},\cdots ,\mathbf{h}_{p}^{n},\cdots ,\mathbf{h}_{p}^{N} \right] \in \mathbb{C} ^{L\times N}$
denotes the channel coefficient matrix between the transmitter and the receiver corresponding to the $p$th frequency offset in the pool $\boldsymbol{P}$ \cite{bacsar2013orthogonal,jian2023mimo}, where {\small $\mathbf{h}_{p}^{n}=\left[ h_{N( p-1) +n}^{1},h_{N( p-1) +n}^{2},\cdots ,h_{N( p-1) +n}^{L} \right] ^T$}. 
For example, when the $p$th frequency offset is assigned to the $n$th antenna, the receive channel vector corresponding to that antenna is the $n$th column of $\mathbf{H}_p$. 
Note that each entry in $\mathbf{H}$ satisfy the independently and identically distributed (i.i.d) $\mathcal{C} \mathcal{N} (0,\sigma ^{2})$.

The system communication sub-function treats each pulse in the same way at the user. In this subsection, for convenience, the first pulse is taken as an example for analysis. Assuming that $\mathbf{f}^{u,v}=[\Delta f_{1}^{u,v},\cdots ,\Delta f_{n}^{u,v},\cdots ,\Delta f_{N}^{u,v}]^T$ is the frequency offset permutation vector transmitted by the BS, the channel matrix involved can be expressed as $\mathring{\mathbf{H}}=\mathbf{HS}=\left[ \mathring{\mathbf{h}}_1,\cdots ,\mathring{\mathbf{h}}_n,\cdots ,\mathring{\mathbf{h}}_N \right] \in \mathbb{C} ^{L\times N}$
with $\mathbf{S}=\left[ \mathbf{s}_1,\cdots ,\mathbf{s}_n,\cdots ,\mathbf{s}_N \right] \in \mathbb{C} ^{NP\times N}$ representing the selection matrix corresponding to the frequency offset permutation vector $\mathbf{f}^{u,v}$. 
Note that only the $(n+N\Delta f_{n}^{u,v}/\Delta f)$th element in $\mathbf{s}_n=[ 0,\cdots ,1,\cdots ,0 ] ^T$ is non-zero. $\mathring{\mathbf{h}}_n=[ \mathring{h}_{n}^{1},\cdots ,\mathring{h}_{n}^{l},\cdots ,\mathring{h}_{n}^{L} ] ^T$ denotes the channel vector between the $n$th transmit antenna and $L$ receive antennas.

Therefore, the signal received by the $l$th receive antenna of the communication user can be expressed as
\begin{equation}\label{ap_yl}
	\begin{aligned}
		y_l(t) &=\sqrt{\frac{P_S}{N}}\sum_{n=1}^N{\phi (t-\tau _{l,n}) e^{j2\pi (f_c+\Delta f_{n}^{u,v}) (t-\tau _{l,n})}\mathring{h}_{n}^{l}x_n}
		\\
		&\approx \phi (t-\tau _C) \sqrt{\frac{P_S}{N}}\sum_{n=1}^N{e^{j2\pi (f_c+\Delta f_{n}^{u,v} ) (t-\tau _l)}\mathring{h}_{n}^{l}x_n}	,
	\end{aligned}
\end{equation}
where $x_n$ denotes the constellation symbol transmitted by the $n$th antenna. $\mathring{h}_{n}^{l}$ and $\tau _{l,n}$ represent the channel cofficient and delay between the $n$th emit antenna and the $l$th receive antenna, respectively. $d_1$, $d_2$ stands for the spacing between neighboring transmit and receive antennas, and $c$ is the speed of light. In \eqref{ap_yl}, the small phase deviation caused by the receive spacing can be offset by a priori knowledge and thus can be neglected \cite{basit2021fda,jian2023mimo}. Hence, we have $\tau _n=\frac{R_C-\left( n-1 \right) d_1\sin \theta _C}{c}$. Moreover, due to the narrowband assumption, the approximation $\phi \left( t-\tau _{l,n} \right) \approx \phi \left( t-\tau _C \right) $ with $\tau _C=\frac{R_C}{c}$ is considered \cite{lan2020suppression,gui2017coherent}.

\begin{figure}[htp]
	\vspace{-0.4cm}
	\centering
	\subfigure	{\includegraphics[width=0.45\textwidth]{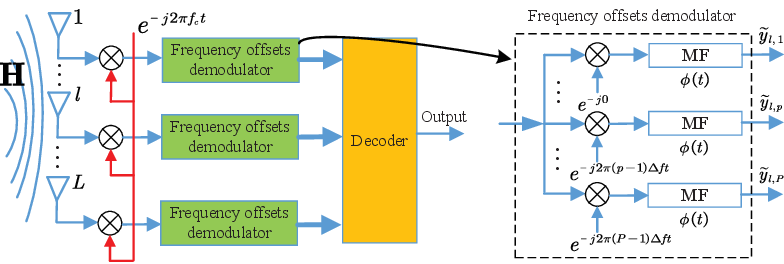}}
	\caption{Communication receiver structure.}
	\vspace{-0.4cm}
	\label{Com_reveiver}
\end{figure}
The communication receiver structure is shown in Fig. \ref{Com_reveiver}. For any receive antenna, the received signal is first down-converted by $e^{-j2\pi f_ct}$ and processed through a frequency offset demodulator with $P$ parallel filters. Specifically, in the $p$th parallel filter, the signal is multipled by $e^{-j2\pi ( p-1) \Delta ft}$ and match-filtered by $\phi \left( t \right)$ at the optimal sampling point of $T_P+\tau _C$. Therefore, referring to \eqref{orthogonality}, the demodulated output signal of the $p$th parallel filter of the $l$th antenna can be formulated as
\begin{equation}\label{y_wave_lp}
	\begin{aligned}
		&\tilde{y}_{l,p}=y_l\left( t \right) \times e^{-j2\pi f_ct}\times e^{-j2\pi \left( p-1 \right) \Delta ft}*\varphi \left( t \right) 
		\\
		&=\left\{ \begin{array}{r}
			\!\!\!\sqrt{\frac{P_S}{N}}e^{-j2\pi \left( f_c+\Delta f_{n}^{u,v} \right) \tau _n}\mathring{h}_{n}^{l}x_n+n_{p}^{l},\Delta f_{n}^{u,v}=\left( p-1 \right) \Delta f\\
			n_{p}^{l},\Delta f_{n}^{u,v}\ne \left( p-1 \right) \Delta f	,\\
		\end{array} \right. 
	\end{aligned}
\end{equation}
where $n_{p}^{l}\sim \mathcal{C} \mathcal{N} \left( 0,\frac{N_0}{P} \right) $ denotes the output additive Gaussian white noise (AWGN) \cite{bacsar2013orthogonal}. $N_0$ is the total output noise power. 

For decoding, traditional ML methods traverse all permutations of frequency offsets and constellation symbols \cite{ji2019physical,nusenu2020space,jian2023physical}, which suffer from high computational complexity. Inspecting \eqref{y_wave_lp} gives that there exist $N$ non-near-to-zero outputs from the frequency offset demodulator, which can be used to estimate the transmitted frequency offsets. Leveraging this observation, an MLTSD method is proposed to reduce the decoding complexity and BER in this paper. 

In the first stage, we stack the $p$th outputs of all $L$ receive antennas into a vector yields 
\begin{equation}\label{y_p_wave}
	\begin{aligned}
		&\tilde{\mathbf{y}}_p=[ \tilde{y}_{1,p},\cdots ,\tilde{y}_{l,p},\cdots ,\tilde{y}_{L,p} ] ^T
		\\
		&=\left\{ \begin{array}{r}
			\!\!\!	\sqrt{\frac{P_S}{N}}e^{-j2\pi ( f_c+\Delta f_{n}^{u,v}) \tau _n}\mathring{\mathbf{h}}_nx_n+\mathbf{n}_p,\Delta f_{n}^{u,v}=(p-1) \Delta f\\
			\!\!\!	\mathbf{n}_p,\Delta f_{n}^{u,v}\ne (p-1) \Delta f  ,\\
		\end{array} \right. 
	\end{aligned}
\end{equation}
where $\mathbf{n}_p=[ n_{p}^{1},\cdots ,n_{p}^{l},\cdots ,n_{p}^{L} ] ^T$ represents the noise vector of the $p$th outputs of $L$ antennas. 

From \eqref{y_p_wave}, the set $\left\{ \tilde{\mathbf{y}}_1,\cdots ,\tilde{\mathbf{y}}_p,\cdots ,\tilde{\mathbf{y}}_P \right\}$ can be obtained. Then, the frequency offsets indices are located by finding $N$ maximum values in the set $\left\{ \left\| \tilde{\mathbf{y}}_1 \right\| ^2,\cdots,\left\| \tilde{\mathbf{y}}_p \right\| ^2,\cdots ,\left\| \tilde{\mathbf{y}}_P \right\| ^2 \right\} $, as \cite{basit2021fda,jian2023mimo}
\begin{equation}\label{fre_est}
	\begin{aligned}
		\left\{ \begin{array}{l}
			\!\!I_1=\underset{I_1}{\mathrm{arg}\max}\left\{ \left\| \tilde{\mathbf{y}}_1 \right\| ^2,\cdots,\left\| \tilde{\mathbf{y}}_p \right\| ^2,\cdots ,\left\| \tilde{\mathbf{y}}_P \right\| ^2 \right\}\\
			\vdots\\
			\!\!I_n=\underset{I_n,I_n\ne I_{n-1}\ne \cdots \ne I_1}{\mathrm{arg}\max}\left\{ \left\| \tilde{\mathbf{y}}_1 \right\| ^2,\cdots,\left\| \tilde{\mathbf{y}}_p \right\| ^2,\cdots ,\left\| \tilde{\mathbf{y}}_P \right\| ^2 \right\}\\
			\vdots\\
			\!\!I_N=\underset{I_N,I_N\ne I_{N-1}\ne \cdots \ne I_1}{\mathrm{arg}\max}\left\{ \left\| \tilde{\mathbf{y}}_1 \right\| ^2,\cdots,\left\| \tilde{\mathbf{y}}_p \right\| ^2,\cdots ,\left\| \tilde{\mathbf{y}}_P \right\| ^2 \right\}	.\\
		\end{array} \right.
	\end{aligned}
\end{equation}
Therefore, the transmit frequency offsets are estimated as $\left\{ \left[ I_1,I_2,\cdots ,I_N \right] -1 \right\} \Delta f$. Note that the estimated indices are not permutated in the first stage, but the permutation and constellation symbols are further estimated in the second stage.

\begin{figure*}[tbp]
	\vspace{-0.4cm}
	\begin{equation}\label{y_In_wave}
		\begin{aligned}
			\tilde{\mathbf{y}}_{I_n}&=\left[ \tilde{y}_{1,I_n},\cdots ,\tilde{y}_{l,I_n},\cdots ,\tilde{y}_{L,I_n} \right] ^T
			\\
			&=\left\{ \begin{array}{r}
				\sqrt{\frac{P_S}{N}}e^{-j2\pi \left( f_c+\left( I_n-1 \right) \Delta f \right) \tau _{\mathring{n}}}\mathbf{h}_{N\left( I_n-1 \right) +\mathring{n}}x_{I_n}+\mathbf{n}_{I_n},\Delta f_{n}^{u,v}=\left( I_n-1 \right) \Delta f\\
				\mathbf{n}_{I_n},\Delta f_{n}^{u,v}\ne \left( I_n-1 \right) \Delta f		,\\
			\end{array} \right.  
		\end{aligned}
	\end{equation}
\vspace{-0.3cm}
\end{figure*}
In the second stage, the $L$ outputs corresponding to the estimated index $I_n$ in \eqref{fre_est} are stacked into a vector as \eqref{y_In_wave}, shown at the top of next page. In \eqref{y_In_wave}, $\mathbf{h}_{N\left( I_n-1 \right) +\mathring{n}}=\left[ h_{N\left( I_n-1 \right) +\mathring{n}}^{1},\cdots ,h_{N\left( I_n-1 \right) +\mathring{n}}^{l},\cdots ,h_{N\left( I_n-1 \right) +\mathring{n}}^{L} \right] \in \mathbf{H}_{I_n}$ and $\mathring{n}$ denote the receive channel vector and antenna index corresponding to $(I_n-1) \Delta f$, respectively. $x_{I_n}$ represents the constellation symbol attached to $(I_n-1) \Delta f$. $\mathbf{n}_{I_n}=[n_{1,I_n},n_{2,I_n},\cdots ,n_{L,I_n}] ^T\sim \mathcal{C} \mathcal{N} (0,\frac{N_0}{P}\mathbf{I}_L) $ stands for the noise vector of the $I_n$th outputs of all $L$ antennas.

Next, the antenna index with transmit frequency offset $(I_n-1) \Delta f$ and the transmitted QAM symbol are estimated by the ML criterion: 
\begin{equation}\label{FOPIM_ML}
	\begin{aligned}
		[ \mathbf{h}_{N(I_n-1) +\hat{n}},\hat{x}_{I_n} ] =&\mathop {arg\min} \limits_{\mathbf{h}_{N\left( I_n-1 \right) +\mathring{n}}\in \mathbf{H}_{I_n},x_{I_n}\in \mathcal{J}}\left\| \tilde{\mathbf{y}}_{I_n}-\sqrt{\frac{P_S}{N}}\times \right. 
		\\
		&\left. e^{-j2\pi ( f_c+(I_n-1) \Delta f ) \tau _{\mathring{n}}}\mathbf{h}_{N( I_n-1 ) +\mathring{n}}x_{I_n} \right\| ^2	,
	\end{aligned}
\end{equation}
where $\hat{n}$ and $\hat{x}_{I_n}$ stand for the estimations of $\mathring{n}$, $x_{I_n}$, respectively. $\mathcal{J}$ is the set of $J$-ary QAM symbols. $\mathbf{H}_{I_n}$ stands for the $I_n$th sub-matrix in $\mathbf{H}$. The calculation of \eqref{y_In_wave} and \eqref{FOPIM_ML} is looped until the transmit antenna indices and constellation symbols corresponding to the frequency offset vector $([I_1,I_2,\cdots ,I_N] -1) \Delta f$ are estimated.

The proposed MLTSD method utilizes the multi-parallel filter structure of the receiver, which is both efficient and simple. Compared with the traditional ML method (seen in the Appendix \ref{ApxA}), which requires $ 2^{\lfloor \log _2C_{P}^{N} \rfloor}\times 2^{\lfloor \log _2N! \rfloor}$ searches \cite{ji2019physical,nusenu2020space,jian2023physical}, the proposed MLTSD method requires only $J\times N^2$ searches, which dramatically reduces the decoding complexity.

\subsection{System sensing sub-function}
We consider an interested target located at any far-field point of $(R_T,\theta _T)$, which is stationary similarly to \cite{liu2020joint,chen2022generalized,lan2020suppression,xu2015range}. A coherent processing interval (CPI) of the radar comprises $K$ pulses. Calling back \eqref{s_kn}, within the $k$th pulse, the received signal of the $m$th element can be written as
\begin{equation}
	\begin{aligned}
		y_{k}^{m}\left( t \right) &=\sqrt{\frac{P_S}{N}}\sum_{n=1}^N{\phi _k\left( t-\tau _{m,n} \right) \xi e^{j2\pi \left( f_c+\Delta f_{k,n}^{u,v} \right) \left( t-\tau _{m,n} \right)}x_{n}^{k}}
		\\
		&\approx \sqrt{\frac{P_S}{N}}\sum_{n=1}^N{\phi _k\left( t-\tau _T \right) \xi e^{j2\pi \left( f_c+\Delta f_{k,n}^{u,v} \right) \left( t-\tau _{m,n} \right)}x_{n}^{k}}		,
	\end{aligned}
\end{equation}
where $\phi _k\left( t-\tau _{m,n} \right) =\phi \left( t-\left( k-1 \right) T-\tau _{m,n} \right)$, $k=1,\cdots ,K$. $\xi$ represents the complex reflection coefficient of the target. {\small $\tau _{m,n}=\frac{2R_T-(n-1) d_1\sin \theta _D-( m-1) d_3\sin \theta _D}{c}$} stands for the round-trip time between the $n$th transmit element and $m$th receive element, where $d_3$ denote the interval between adjacent antennas of the receiver, respectively. Note that the approximation  $\phi _k\left( t-\tau _{m,n} \right) \approx \phi _k\left( t-\tau _T \right)$ is considered due to the narrow-band consumption, where $\tau _0=2R_T/c$ \cite{lan2020suppression,gui2017coherent}. 

\begin{figure}[htp]
	\vspace{-0.4cm}
	\centering
	\subfigure	{\includegraphics[width=0.45\textwidth]{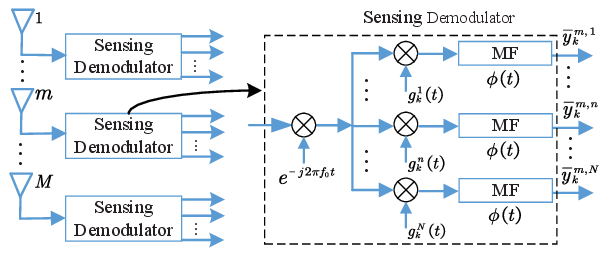}}
	\caption{Sensing receiver structure.}
	\vspace{-0.2cm}
	\label{Sen_reveiver}
\end{figure}
The receiver structure for the sensing function is depicted in Fig. \ref{Sen_reveiver}. The received signal is first downconverted with $e^{-j2\pi f_0t}$ and then fed into $N$ parallel filters. In the $n$th parallel filter, the signal is multiplied by $g_{k}^{n}(t)$ with
\begin{equation}\label{g_kn}
	\begin{aligned}
		g_{k}^{n}\left( t \right) =\left( e^{j2\pi \Delta f_{k,n}^{u,v}t}x_{n}^{k} \right) ^{\dagger}		.
	\end{aligned}
\end{equation}
Note that since the sensing transmitter and receiver are co-located, the receiver has the sufficient priori knowledge to realize \eqref{g_kn} \cite{sturm2011waveform,ma2021frac}. Next, the multiplied signal is match-filtered by $\phi ( t)$. Therefore, the $n$th demodulated output of the $m$th receive antenna can be expressed as
\begin{equation}\label{y_k_mn}
	\begin{aligned}
		&\bar{y}_{k}^{m,n}=y_{k}^{m}( t) e^{-j2\pi f_0t}g_{k}^{n}( t) *\phi ( t)
		\\
		&\approx \sqrt{\frac{P_S}{N}}\xi e^{-j2\pi \frac{2f_cR_T+2\Delta f_{k,n}^{u,v}R_T-f_c( n-1) d_1\sin \theta _T-f_c(m-1) d_3\sin \theta _T}{c}}		.
	\end{aligned}
\end{equation}
Note that the $e^{j2\pi \frac{\Delta f_{k,n}^{u,v}(n-1) d_1\sin \theta _T}{c}}$, $e^{j2\pi \frac{\Delta f_{k,n}^{u,v}(m-1) d_3\sin \theta _T}{c}}$ terms in \eqref{y_k_mn} are tiny enough to be ignored \cite{lan2020glrt,xiong2017fda}.

Further, the outputs of the $m$th receive antenna are written in vector form as
\begin{equation}\label{y_k_mn_vec}
	\begin{aligned}
		\bar{\mathbf{y}}_{k}^{m}=\sqrt{\frac{P_S}{N}}\xi \boldsymbol{\omega }_ke^{j2\pi \frac{d_3}{\lambda _0}\left( m-1 \right) \sin \theta _T}	,
	\end{aligned}
\end{equation}
where $\boldsymbol{\omega }_k=\left[ e^{-j2\pi \psi _{k,1}},e^{-j2\pi \psi _{k,2}},\cdots ,e^{-j2\pi \psi _{k,N}} \right] ^T$ with $\psi _{k,n}=2\frac{\Delta f_{k,n}^{u,v}}{c}R_T-\frac{d_1}{\lambda _0}\left( n-1 \right) \sin \theta _T$ and $\lambda _0=c/f_c$. The $e^{-j\frac{4\pi f_cR_T}{c}}$ term is absorbed into $\xi$. 

Therefore, the $k$th data snapshot of the received signal can be expressed as
\begin{equation}\label{y_k}
	\begin{aligned}
		\bar{\mathbf{y}}_k&=[ \bar{\mathbf{y}}_{k,1}^{T},\bar{\mathbf{y}}_{k,2}^{T},\cdots ,\bar{\mathbf{y}}_{k,M}^{T} ] ^T+\bar{\mathbf{n}}
		\\
		&=\sqrt{\frac{P_S}{N}}\xi \mathbf{a}\left( \theta _T \right) \otimes \left[ \mathbf{b}_k\left( R_T \right) \odot \mathbf{c}_k\left( \theta _T \right) \right] +\bar{\mathbf{n}}	,
	\end{aligned}
\end{equation}
where $\bar{\mathbf{n}}\sim \mathcal{C} \mathcal{N} ( 0,N_1\boldsymbol{I}_{MN} )$ represents received noise. $\mathbf{a}(\theta _T)$, $\mathbf{b}_k\left( R_T \right)$ and $\mathbf{c}\left( \theta _T \right)$ stand for the receive steering vector, the transmit range steering vector and the transmit angle steering vector, respectively, which are given as
\begin{equation}\label{a_T}
	\begin{aligned}
		\mathbf{a}\left( \theta _T \right) =[ 1,e^{j2\pi \frac{d_3}{\lambda _0}\sin \theta _T},\cdots ,e^{j2\pi \frac{d_3}{\lambda _0}\left( M-1 \right) \sin \theta _T} ] ^T	,
	\end{aligned}
\end{equation}
\begin{equation}\label{b_T}
	\begin{aligned}
		\mathbf{b}_k(R_T) \!=\![ e^{-j4\pi \frac{\Delta f_{k,1}^{u,v}}{c}R_T},e^{-j4\pi \frac{\Delta f_{k,2}^{u,v}}{c}R_T},\!\cdots \!,e^{-j4\pi \frac{\Delta f_{k,N}^{u,v}}{c}R_T} ] ^T	,
	\end{aligned}
\end{equation}
and
\begin{equation}\label{c_T}
	\begin{aligned}
		\mathbf{c}\left( \theta _T \right) =[ 1,e^{j2\pi \frac{d_1}{\lambda _0}\sin \theta _T},\cdots ,e^{j2\pi \frac{d_1}{\lambda _0}\left( N-1 \right) \sin \theta _T} ] ^T	,
	\end{aligned}
\end{equation}
respectively. From \eqref{b_T}, one can observe that the transmit steering vector of FDA-MIMO is range dependent, which can be used to improve the distance estimation accuracy \cite{gui2017coherent,xiong2017fda,xu2015joint}.

When solely utilizing the traditional pulse compression method, FDA-MIMO's distance estimation accuracy identical to MIMO radar, remains at $c/2B$, where $B=\Delta f$ denotes the transmit signal bandwidth \cite{gui2017coherent,lan2020glrt,xu2015joint}. Take this into mind, we propose employing the TSMLE for accurate target distance estimation. Specifically, fast-time matched filtering yields coarse range estimation $r_{\mathrm{c}}$, and we introduce the compensation vector for $k$th pulse at receiver as
\begin{equation}
	\begin{aligned}
		\mathbf{w}_k (r_{\mathrm{c}})=\mathbf{1}_M\otimes \mathbf{d}(r_{\mathrm{c}})	,
	\end{aligned}
\end{equation}
where $\mathbf{1}_M$ denotes one vector, and
\begin{equation}
	\begin{aligned}
		\mathbf{d}_k\left( r_{\mathrm{c}} \right) =[ 1,e^{j4\pi r_{\mathrm{c}}\Delta f_{k,1}^{u,v}/c},\cdots ,e^{-j4\pi r_{\mathrm{c}}\Delta f_{k,N}^{u,v}/c} ] ^T	.
	\end{aligned}
\end{equation}

After compensation, the received signal in \eqref{y_k} can be rewritten as
\begin{equation}\label{y_k_bf}
	\begin{aligned}
		\bar{\mathbf{y}}_{k}^{\mathrm{comp}}&=\mathbf{w}_k\left( r_{\mathrm{c}} \right) \odot \tilde{\mathbf{y}}_k
		\\
		&=\xi \mathbf{v}_k\left( \theta _T,q,\Delta r \right) +\bar{\mathbf{n}}_k	,
	\end{aligned}
\end{equation}
where $\bar{\mathbf{n}}_k=\mathbf{w}_k\left( r_{\mathrm{c}} \right) \odot \bar{\mathbf{n}}$ and
\begin{equation}
	\begin{aligned}
		\mathbf{v}_k\left( \theta _T,q,\Delta r \right) =\sqrt{\frac{P_S}{N}}\mathbf{a}\left( \theta _T \right) \otimes \left[ \mathbf{b}_k\left( \Delta r \right) \odot \mathbf{c}_k\left( \theta _T \right) \right] 	,
	\end{aligned}
\end{equation}
with $\Delta r=R_T-r_{\mathrm{c}}$ denotes the principal distance difference and $\Delta r\in [ -\frac{c}{4B},\frac{c}{4B} ]$.  

In the first step, the compensated received signal is reconstructed into an $M\times N$ dimensional signal matrix as
\begin{equation}
	\begin{aligned}
		\bar{\mathbf{Y}}_{k}^{\mathrm{comp}}=\sqrt{\frac{P_S}{N}}\xi \mathbf{a}\left( \theta _T \right) \left[ \mathbf{b}_k\left( \Delta r \right) \odot \mathbf{c}_k\left( \theta _T \right) \right] ^T+\bar{\mathbf{N}}_k	,
	\end{aligned}
\end{equation}
where $\bar{\mathbf{N}}_k\in \mathbb{C} ^{M\times N}$ is yielded from the noise vector $\bar{\mathbf{n}}_k$ matrixed. The $n$th column of $\bar{\mathbf{Y}}_{k}^{\mathrm{comp}}$ represents the received data corresponding to the $n$th transmit antenna. Then, the target's angle is estimated by \footnote{Note that to reduce the computation, the search range can be reduced by using a coarse angle estimation (e.g., calculate the DFT on the received array signal as in ref. \cite{xu2023bandwidth}), and then refined by using \eqref{theta_T}.}
\begin{equation}\label{theta_T}
	\begin{aligned}
		\hat{\theta}_T=\mathrm{arg}\max_{\theta _T\in \left[ -90\degree,90\degree \right]} \left\| \mathbf{a}\left( \theta _T \right) ^H\bar{\mathbf{Y}}_{k}^{\mathrm{comp}} \right\| ^2	,
	\end{aligned}
\end{equation}
where $\mathbf{a}(\theta _T)$ the receive beamformer, $\hat{\theta}_T$ stands for the estimation of $\theta _T$.

In the second step, we use the maximum likelihood criterion to estimate the target distance. Taking the results $\hat{\theta}_T$ of \eqref{theta_T} into \eqref{y_k_bf} yields the likelihood function as
\begin{equation}
	\begin{aligned}
		&f(\theta _T,\Delta r,\beta ;\bar{\mathbf{y}}_{k}^{\mathrm{comp}})
		\\
		&= \frac{1}{\pi ^{MN} \det ( \mathbf{Q} )} e^{- [ \bar{\mathbf{y}}_{k}^{\mathrm{comp}}\!-\!\beta \mathbf{v}_k( \hat{\theta}_T,\Delta r ) ] ^{\dagger}\mathbf{Q}^{-1}[ \bar{\mathbf{y}}_{k}^{\mathrm{comp}}\!-\!\beta \mathbf{v}_k( \hat{\theta}_T,\Delta r ) ] }	,
	\end{aligned}
\end{equation}
where $\mathbf{Q}=\sigma _{1}^{2}\boldsymbol{I}_{M\times N}$ denotes the noise covariance matrix.

Maximizing the likelihood function is equivalent to minimizing the following 
\begin{equation}\label{de_r_hat}
	\begin{aligned}
		\left[ \Delta \hat{r} \right] =\underset{\Delta r\in [-\frac{c}{4B},\frac{c}{4B}]}{arg\min}
		&[ \bar{\mathbf{y}}_{k}^{\mathrm{comp}}-\beta \mathbf{v}_k(\hat{\theta}_T,\Delta r) ] ^{\dagger}\times 
		\\
		&\mathbf{Q}^{-1}[ \bar{\mathbf{y}}_{k}^{\mathrm{comp}}-\beta \mathbf{v}_k( \hat{\theta}_T,\Delta r ) ] 	, 
	\end{aligned}
\end{equation}
where $\Delta \hat{r}$ denotes the eatimation of $\Delta r$.

Derivative operation on \eqref{de_r_hat} yields the optimal $\beta$ value as
\begin{equation}\label{beta_hat}
	\begin{aligned}
		\hat{\beta}=\frac{\mathbf{v}_k(\hat{\theta}_T,\Delta r) ^H\mathbf{Q}^{-1}\bar{\mathbf{y}}_{k}^{\mathrm{comp}}}{\mathbf{v}_k(\hat{\theta}_T,\Delta r) ^H\mathbf{Q}^{-1}\mathbf{v}_k(\hat{\theta}_T,\Delta r)}		.
	\end{aligned} 
\end{equation}

Carrying \eqref{beta_hat} into \eqref{de_r_hat} yields
\begin{equation}\label{max_MLE}
	\begin{aligned}
		\left[ \Delta \hat{r} \right] =\underset{\Delta r\in \left[ -\frac{c}{4B},\frac{c}{4B} \right]}{arg\max}\frac{\left| ({\bar{\mathbf{y}}_{k}^{\mathrm{comp}}})^H\mathbf{Q}^{-1}\mathbf{v}_k( \hat{\theta}_T,\Delta r ) \right|^2}{\mathbf{v}_k( \hat{\theta}_T,\Delta r ) ^H\mathbf{Q}^{-1}\mathbf{v}_k( \hat{\theta}_T,\Delta r )}	. 
	\end{aligned}
\end{equation}
That is, the optimal distance $\Delta \hat{r}$ is estimated by searching in $[-\frac{c}{4B},\frac{c}{4B}]$. Examining \eqref{max_MLE} shows that FDA-MIMO can perform a secondary search within a range cell, achieving super-resolution in range dimension \cite{gui2017coherent,lan2020glrt,xu2015joint}.

\section{System communication and sensing performance analysis}
\label{sc4}
In this section, we analyze system communication and sensing performances separately. We derive the tight upper bound on the system BER for communication, while for sensing, we deduce the CRB closed-form expression.

\subsection{System BER performance analysis}
The information bits carried by the proposed FOPIM method are divided into index bits $k_{\mathrm{IM}}$ and constellation bits $k_{\mathrm{con}}$. Encouraged by this fact, the average bit error probability (ABEP) of the proposed MLTSD method is formulated as
\begin{equation}\label{P_MLTSD}
	\begin{aligned}
		P_{\mathrm{MLTSD}}=\frac{P_{\mathrm{IM}}k_{\mathrm{IM}}+P_{\mathrm{con}}k_{\mathrm{con}}}{k_{\mathrm{IM}}+k_{\mathrm{con}}}	,
	\end{aligned} 
\end{equation}
where $P_{\mathrm{IM}}$ and $P_{\mathrm{con}}$ stand for the ABER of index bits and constellation bits, respectively.

We first deduce $P_{\mathrm{IM}}$. One can find that there are totally $C_{k_{\mathrm{IM}}}^{r}$ events by erring $r\in \left[ 1,\cdots ,k_{\mathrm{IM}} \right]$ bits out of $k_{\mathrm{IM}}$ bits. Without loss of generality, we consider that once an error occurs, there is an equal probability that the emission frequency offset permutation is incorrectly estimated to be any of the remaining ones. Therefore, the ABEP of the index bits can be modeled as
\begin{equation}\label{P_IM}
	\begin{aligned}
		P_{\mathrm{IM}}&=\frac{P_{\Delta f}}{\left( 2^{k_{\mathrm{IM}}}-1 \right) k_{\mathrm{IM}}}\sum_{r=1}^{k_{\mathrm{IM}}}{rC_{k_{\mathrm{IM}}}^{r}}
		\\
		&=\frac{2^{k_{\mathrm{IM}}}P_{\Delta f}}{2\left( 2^{k_{\mathrm{IM}}}-1 \right)}	, 
	\end{aligned}
\end{equation}
where $P_{\Delta f}$ denotes the probability that the selected frequency offset permutation is incorrectly detected, which can be calculated as
\begin{equation}
	\begin{aligned}
		P_{\Delta f}=1-\left( 1-P_{\mathrm{comb}} \right) \left( 1-P_{\mathrm{perm}} \right) , 
	\end{aligned}
\end{equation}
where $P_{\mathrm{comb}}$ and $P_{\mathrm{perm}}$ represent probabilities of incorrect first-stage frequency offset estimation and second-stage frequency offset permutation misalignment, respectively.

To derive $P_{\mathrm{comb}}$, we begin by calculating the probability $P_e$ that any output containing the signal is smaller than any output containing only noise. Calling back \eqref{fre_est}, assuming that of $P$ filter outputs, the $p$th filter outputs contain signal and the $p'$th contain only noise, i.e. $\tilde{\mathbf{y}}_p=\sqrt{\frac{P\mathrm{s}}{N}}\mathring{\mathbf{h}}_{n}^{}e^{-j2\pi ( f_c+\Delta f_{n}^{u,v} ) \tau _n}x_n+\mathbf{n}_p$ and $\tilde{\mathbf{y}}_{p'}=\mathbf{n}_{p'}\sim \mathcal{C} \mathcal{N} (0,\frac{N_0}{P}\mathbf{I}_L) $. Then, we have
\begin{equation}\label{P_e_or}
	\begin{aligned}
		P_e=P\left( \left\| \tilde{\mathbf{y}}_{p\prime} \right\| ^2-\left\| \tilde{\mathbf{y}}_p \right\| ^2>0 \right) . 
	\end{aligned}
\end{equation}

The elements of $\tilde{\mathbf{y}}_p$ are i.i.d random variables, following $\mathcal{C} \mathcal{N} \left( 0,\frac{P\mathrm{s}}{N}\sigma ^2+\frac{N_0}{P} \right)$. Moreover, each element in $\tilde{\mathbf{y}}_{p\prime}$ follows i.i.d  $\mathcal{C} \mathcal{N} \left( 0,\frac{N_0}{P} \right) $. Hence, $\left\| \tilde{\mathbf{y}}_p \right\| ^2$ and $\left\| \tilde{\mathbf{y}}_{p'} \right\| ^2$ satisfy chi-squared distributions with $2L$ degrees of freedom (DoF) with probability density functions (PDFs) of
\begin{equation}
	\begin{aligned}
		f_{\left\| \tilde{\mathbf{y}}_p \right\| ^2}\left( x \right) =\frac{1}{2^L\Gamma \left( L \right) \left( \sigma _{1}^{2} \right) ^L}x^{L-1}\exp \left( -\frac{x}{2\sigma _{1}^{2}} \right) 	, 
	\end{aligned}
\end{equation}
and
\begin{equation}
	\begin{aligned}
		f_{\left\| \tilde{\mathbf{y}}_{p'} \right\| ^2}\left( x \right) =\frac{1}{2^L\Gamma \left( L \right) \left( \sigma _{2}^{2} \right) ^L}x^{L-1}\exp \left( -\frac{x}{2\sigma _{2}^{2}} \right) 	,
	\end{aligned} 
\end{equation}
respectively, where $\sigma _{1}^{2}=\frac{P\mathrm{s}}{2N}\sigma ^2\left| x_{n} \right|^2+\frac{N_0}{2P}$ and $\sigma _{2}^{2}=\frac{N_0}{2P}$.

Let $Y=\left\| \tilde{\mathbf{y}}_{p\prime} \right\| ^2-\left\| \tilde{\mathbf{y}}_p \right\| ^2$, the PDF of $Y$ can be written as \cite{simon2002probability}
\begin{equation}\label{pdf_Y}
	\begin{aligned}
		&p_Y( y|x_n ) =
		\frac{1}{2\sigma _{2}^{2}}\exp \left( -\frac{y}{2\sigma _{2}^{2}} \right) \frac{1}{(L-1)!}\left( \frac{\sigma _{2}^{2}}{\sigma _{2}^{2}+\sigma _{1}^{2}} \right) ^L
		\\
		&\times \sum_{i=0}^{L-1}{\frac{( 2(m-1)-i ) !}{i!(m-1-i)!}}\left( \frac{\sigma _{1}^{2}}{\sigma _{2}^{2}+\sigma _{1}^{2}} \right) ^{L-1-i}\left( \frac{y}{2\sigma _{2}^{2}} \right) ^i, \mathrm{if} y\ge 0	.	
	\end{aligned} 
\end{equation}
Note that $x_n$ is embeded in $\sigma _{1}^{2}$.

Further, the conditional probability of $Y>0$ can be calculated as
\begin{equation}\label{P_Y_bigger_0}
	\begin{aligned}
		P\left( Y>0|x_n \right) 
		=&\frac{1}{2\sigma _{2}^{2}}\frac{1}{(L-1)!}\left( \frac{\sigma _{2}^{2}}{\sigma _{2}^{2}+\sigma _{1}^{2}} \right) ^L\times
		\\
		& \sum_{i=0}^{L-1}{\frac{\left( 2(L-1)-i \right) !}{i!(L-1-i)!}\left( \frac{\sigma _{1}^{2}}{\sigma _{2}^{2}+\sigma _{1}^{2}} \right) ^{L-1-i}i!2\sigma _{2}^{2}}		.
	\end{aligned} 
\end{equation}

By averaging \eqref{P_Y_bigger_0} over $x_n$, \eqref{P_e_or} can be rewritten as
\begin{equation}\label{P_e}
	\begin{aligned}
		P_e=&\frac{1}{J}\sum_{j=0}^J{P\left( Y>0|x_n=c_j \right)}
		\\
		=&\frac{1}{J}\sum_{j=0}^J{\left\{ \frac{1}{2\sigma _{2}^{2}}\frac{1}{(L-1)!}\left( \frac{\sigma _{2}^{2}}{\sigma _{2}^{2}+\sigma _{1}^{2}} \right) ^L \right.} \times 
		\\
		&\left. \sum_{i=0}^{L-1}{\frac{\left( 2(L-1)-i \right) !}{i!(L-1-i)!}\left( \frac{\sigma _{1}^{2}}{\sigma _{2}^{2}+\sigma _{1}^{2}} \right) ^{L-1-i}i!2\sigma _{2}^{2}} \right\}	,
	\end{aligned} 
\end{equation}
where $c_j$ represents the $j$th element in the $J$-ary QAM constellation set $\mathcal{J} $. 

Consequently, the upper bound on the probability of incorrect detection for the emitted frequency offsets is derived as
\begin{equation}\label{P_comb}
	\begin{aligned}
		P_{\mathrm{comb}}\leqslant 1-\left[(1-P_e) ^{P-N}  \right] ^N	. 
	\end{aligned}
\end{equation}

Subsequently, we deduce $P_{\mathrm{perm}}$. Referring to \eqref{FOPIM_ML}, we find that the antenna index and constellation symbol of the frequency offset are jointly detected, while the signal detection among individual transmitting antennas is independent. In the case of the transmit frequency offset being misestimated in the first stage, we consider that its corresponding transmit antenna index and constellation symbol are also misestimated in the second stage. Then, we have
\begin{equation}\label{P_perm}
	\begin{aligned}
		P_{\mathrm{perm}}=1-\left( 1-P_{\mathcal{P}} \right) ^N		,
	\end{aligned} 
\end{equation}
where $P_{\mathcal{P}}$ denotes the probability that the antenna index of a certain frequency offset is incorrectly detected, which is formulated by
\begin{equation}\label{P_P}
	\begin{aligned}
		P_{\mathcal{P}}=\frac{N-1}{N}P_O+\left( 1-P_O \right) P_I	, 
	\end{aligned}
\end{equation}
where $P_O=1-\left( 1-P_e \right) ^{P-N}$ denotes the probability that a given frequency offset is mistakenly estimated in the first stage. $P_I$ represents the probability that the transmit frequency offset is correctly estimated and its corresponding transmit antenna index is incorrectly estimated.

To derive $P_I$, we begin with calculating the conditional pairwise error probability (PEP) that $\mathbf{h}_{N\left( I_n-1 \right) +\mathring{n}}$ is erroneously detected as $\mathbf{h}_{N\left( I_n-1 \right) +\hat{n}}$ on $\mathbf{H}_{I_n}$, i.e.
\begin{equation}\label{P_r_orig}
	\begin{aligned}
		&\mathrm{Pr}\left( \mathbf{h}_{N\left( I_n-1 \right) +\mathring{n}}\rightarrow \mathbf{h}_{N\left( I_n-1 \right) +\hat{n}} \middle| \mathbf{H}_{I_n} \right) 
		\\
		&=\mathrm{Pr}\!\left(\!\! \begin{array}{c}
			\!\!\!\!\left\| \tilde{\mathbf{y}}_{I_n}\!-\!\sqrt{\frac{P_S}{N}}e^{-j2\pi (f_c+(I_n\!-\!1) \Delta f) \tau _{\mathring{n}}}\mathbf{h}_{N(I_n\!-\!1) +\mathring{n}}x_{I_n} \right\| ^2\\
			\!\!\!>\!\!\left\| \tilde{\mathbf{y}}_{I_n}\!\!-\!\sqrt{\frac{P_S}{N}}e^{-j2\pi (f_c+ (I_n\!-\!1) \Delta f) \tau _{\hat{n}}}\mathbf{h}_{N (I_n\!-\!1) +\hat{n}}\hat{x}_{I_n} \right\| ^2\\
		\end{array} \!\!\!\!\right) 
		\\
		&=Q\left( \sqrt{\frac{PP_S\varLambda}{2NN_0}} \right) 	,
	\end{aligned} 
\end{equation}
where 
\begin{equation}
	\begin{aligned}
		\varLambda =\left\| \mathrm{Re}\left\{ \mathfrak{a} -\mathfrak{b} \right\} +j\mathrm{Im}\left\{ \mathfrak{a} -\mathfrak{b} \right\} \right\| ^2	, 
	\end{aligned}
\end{equation}
with
\begin{equation}\label{a_b}
	\begin{aligned}
		\left\{ \begin{array}{c}
			\mathfrak{a} =e^{-j2\pi \left( f_c+\left( I_n-1 \right) \Delta f \right) \tau _{\hat{n}}}\mathbf{h}_{N\left( I_n-1 \right) +\hat{n}}\hat{x}_{I_n}\\
			\mathfrak{b} =e^{-j2\pi \left( f_c+\left( I_n-1 \right) \Delta f \right) \tau _{\mathring{n}}}\mathbf{h}_{N\left( I_n-1 \right) +\mathring{n}}x_{I_n} .\\
		\end{array} \right. 
	\end{aligned}
\end{equation}
From \eqref{a_b}, we can extract that $\varLambda =\sum_{l=1}^{2L}{\psi _{l}^{2}}$, where ${\psi_l}\sim\mathcal{C} \mathcal{N}(0,\sigma _{3}^{2}) $ with
\begin{equation}
	\begin{aligned}
		\sigma _{3}^{2}=\left\{ \begin{array}{c}
			\frac{\sigma ^2\left( \left| \hat{x}_{I_n} \right|^2+\left| x_{I_n} \right|^2 \right)}{2},\hat{n}\ne \mathring{n}\\
			\frac{\sigma ^2\left| \hat{x}_{I_n}-x_{I_n} \right|^2}{2},\hat{n}=\mathring{n}		.\\
		\end{array} \right. 
	\end{aligned} 
\end{equation}

Therefore, $\varLambda$ follows the chi-squared distribution with $2L$ DoF, whose PDF is given 
\begin{equation}\label{f_lam}
	\begin{aligned}
		f_{\varLambda}\left( x \right) =\frac{1}{2^L\Gamma \left( L \right) \left( \sigma _{3}^{2} \right) ^L}x^{L-1}\exp \left( -\frac{x}{2\sigma _{3}^{2}} \right) 	. 
	\end{aligned}
\end{equation}

Averaging \eqref{P_r_orig} on $\varLambda$ gives that 
\begin{equation}\label{Pr_hh}
	\begin{aligned}
		&\mathrm{Pr}\left( \mathbf{h}_{N\left( I_n-1 \right) +\mathring{n}}\rightarrow \mathbf{h}_{N\left( I_n-1 \right) +\hat{n}} \right) 
		\\
		&=\left[ P\left( \alpha \right) \right] ^L\sum_{k=0}^{L-1}{\left( \begin{array}{c}
				L-1+k\\
				k\\
			\end{array} \right) \left[ 1-P\left( \alpha \right) \right] ^k}	, 
	\end{aligned}
\end{equation}
where $P(\alpha )=\frac{1}{2}\left( 1-\sqrt{\frac{\alpha}{1+\alpha}} \right)$ with $\alpha =\frac{PP_S\sigma _{3}^{2}}{2NN_0}$. Consequently, the upper bound of $P_I$ is formulated by
\begin{equation}\label{P_I}
	\begin{aligned}
	P_I\leqslant \frac{1}{NJ}\sum_{\begin{array}{c}
			\mathring{n},\hat{n}\\
			\mathring{n}\ne \hat{n}\\
	\end{array}}{\sum_{x_{I_n},\hat{x}_{I_n}}{\mathrm{Pr}\left( \mathbf{h}_{N\left( I_n-1 \right) +\mathring{n}}\rightarrow \mathbf{h}_{N\left( I_n-1 \right) +\hat{n}} \right)}}	. 
	\end{aligned}
\end{equation}
Finally, the upper bound of $P_{\mathrm{perm}}$ can be obtained by substituting  \eqref{P_I}, \eqref{P_P} into \eqref{P_perm}.

Next, we derive the ABEP of QAM constellation bits, i.e., $P_{\mathrm{con}}$. \eqref{FOPIM_ML} indicates that the QAM symbols of each transmit antenna are detected independently. Therefore, $P_{\mathrm{con}}$ can be treated as the probability of an individual constellation symbol being incorrectly estimated. We consider that the constellation symbol can only be correctly estimated if the frequency offset and its antenna index are correctly estimated. Considering this, we have \cite{jian2023physical}
\begin{equation}\label{P_con}
	\begin{aligned}
		P_{\mathrm{con}}=\frac{1}{2}P_{\mathcal{P}}+\left( 1-P_{\mathcal{P}} \right) P_{\mathrm{QAM}}	, 
	\end{aligned}
\end{equation}
where $P_{\mathrm{QAM}}$ denotes the ABEP of the QAM symbol under the frequency offset index is correctly detected.

This paper considers the more general rectangular QAM modulation, which can be viewed as consisting of two pulse amplitude modulation (PAM) modulations that are orthogonal to each other, i.e., $\mu $-ary PAM of the I-signal and $\eta$-ary PAM of the Q-signal, $J=\mu \times \eta$. According to the results of \cite{cho2002general}, the conditional error probability of the $b$th bit in in-phase components is expressed as
\begin{equation}
	\begin{aligned}
		&P_{\mu -\mathrm{PAM}}\left( b \middle| \gamma _{Tot} \right)
		\\
		&=\frac{2}{\mu}\sum_{i=0}^{\left( 1-2^{-b} \right) I-1}{\left\{ \left( -1 \right) ^{\lfloor \frac{i\cdot 2^{b-1}}{\mu} \rfloor} \right.}\times 
		\\
		&\quad\left( 2^{b-1}-\left. \lfloor \frac{i\cdot 2^{b-1}}{\left. \mu \right.}+\frac{1}{2} \rfloor \right. \right) \left. Q\left( (2i+1)g\sqrt{2\gamma _{Tot}} \right) \right\} , 
	\end{aligned}
\end{equation}
where $g=\sqrt{\frac{3}{\mu ^2+\eta ^2-2}}$ represents the minimum Euclidean distance between two constellation points. $\gamma _{Tot}=\sum_{l=1}^L{\gamma _l}$ stands for the total received signal to noise ratio (SNR) with $\gamma _l=\frac{PP_S\left| h_{N\left( I_n-1 \right) +\mathring{n}}^{l}x_{I_n} \right|^2}{NN_0}$. Therefore, the PDF of $\gamma _{Tot}$ can be written as
\begin{equation}\label{pdf_tot}
	\begin{aligned}
		f_{\gamma _{Tot}}(x)=\frac{1}{2^L\Gamma \left( L \right) \left( \frac{PP_S\sigma ^2}{2NN_0} \right) ^L}x^{L-1}\exp \left( -\frac{NN_0}{PP_S\sigma ^2}x \right)	. 
	\end{aligned}
\end{equation}

By averaging \eqref{pdf_tot} on $\gamma _{Tot}$, we have
\begin{equation}
	\begin{aligned}
		&P_{\mu -\mathrm{PAM}}\left( b \right) 
		\\
		&=\frac{2}{\left. \mu \right.}\sum_{i=0}^{\left( 1-2^{-b} \right) \left. \mu \right. -1}{\left\{ (-1) ^{\lfloor \frac{i\cdot 2^{b-1}}{\left. \mu \right.} \rfloor} \right.} \left( 2^{b-1}\!-\!\left. \lfloor \frac{i\cdot 2^{b-1}}{\left. \mu \right.}+\frac{1}{2} \rfloor \right. \right) 
		\\
		&\quad \times \left. P\left( \beta \right) ^L\sum_{k=0}^{L-1}{\left( \begin{array}{c}
				L-1+k\\
				k\\
			\end{array} \right)}\left[ 1-P\left( \beta \right) \right] ^k \right\} ,  
	\end{aligned}
\end{equation}
where $P\left( \beta \right) =\frac{1}{2}\left( 1-\sqrt{\frac{\beta}{1+\beta}} \right) $ with $\beta =\frac{[ (2i+1)g ] ^2PP_S\sigma ^2}{NN_0}$.

Similarly, the conditional error probability of the $b$-th bits in quadrature components can be written as
\begin{equation}
	\begin{aligned}
		&P_{\eta -\mathrm{PAM}}\left( b \right) 
		\\
		&=\frac{2}{\eta}\sum_{i=0}^{\left( 1-2^{-b} \right) \eta -1}{\left\{ ( -1 ) ^{\lfloor \frac{i\cdot 2^{b-1}}{\eta} \rfloor} \right.} \left( 2^{b-1}-\left. \lfloor \frac{i\cdot 2^{b-1}}{\left. \eta \right.}+\frac{1}{2} \rfloor \right. \right) 
		\\
		&\quad \times \left. P\left( \beta \right) ^L\sum_{k=0}^{L-1}{\left( \begin{array}{c}
				L-1+k\\
				k\\
			\end{array} \right)}\left[ 1-P\left( \beta \right) \right] ^k \right\}  .
	\end{aligned} 
\end{equation}

Further, $P_{\mathrm{QAM}}$ can be formulated as
\begin{equation}\label{P_QAM}
	\begin{aligned}
		P_{\mathrm{QAM}}=\frac{1}{\log _2(\mu \eta)}\left( \sum_{b=1}^{\log _2\mu}{P_{\mu -\mathrm{PAM}}\left( b \right)} \!+\! \sum_{b=1}^{\log _2\eta}{P_{\eta -\mathrm{PAM}}\left( b \right)} \right) 	.
	\end{aligned}
\end{equation}
Finally, substituting \eqref{P_QAM}, \eqref{P_con}, \eqref{P_IM} into \eqref{P_MLTSD} yields the ABEP upper bound for the proposed MLTSD approach.

\subsection{System CRB analysis}
In this section, we study the sensing performance by deriving the CRBs for the range and angle. Calling back \eqref{y_k_bf}, the $k$th snapshot noise-free data is denoted as
\begin{equation}\label{u_k}
	\begin{aligned}
		\boldsymbol{u}_k=\sqrt{\frac{P_{\mathrm{S}}}{N}}\xi \bar{\boldsymbol{u}}_k	,
	\end{aligned}
\end{equation}
where $\bar{\boldsymbol{u}}_k=\mathbf{a}\left( \theta _T \right) \otimes \left[ \mathbf{b}_k\left( \Delta r \right) \odot \mathbf{c}\left( \theta _T \right) \right] $ represents the transmit-receive joint steering vector of the $k$th snapshot.

Thereafter, the estimated parameters are modeled as 
\begin{equation}
	\begin{aligned}
		\boldsymbol{\alpha }=\left[ R_D,\theta _D,\mathrm{Re}\left\{ \xi \right\} ,\mathrm{Im}\left\{ \xi \right\} \right] ^T	.
	\end{aligned}
\end{equation}
Then, the CRBs of the parameters in $\boldsymbol{\alpha }$ are then given as the diagonal elements of $\mathbf{F}^{-1}$, where $\mathbf{F}$ is the Fisher information matrix (FIM) with $K$ snapshots, and its $(i,j)$th element is denoted as
\begin{equation}\label{F_ij}
	\begin{aligned}
		F_{i,j}=\frac{2P_{\mathrm{S}}}{N}\mathrm{Re}\sum_{k=1}^K{\left\{ \frac{\partial \left( \xi \bar{\boldsymbol{u}}_k \right) ^H}{\partial \alpha _i}\mathbf{Q}^{-1}\frac{\partial \xi \bar{\boldsymbol{u}}_k}{\partial \alpha _j} \right\}}	,
	\end{aligned}
\end{equation}
where $\alpha _i$ and $\alpha _j$ stand for the $i$th and $j$th element of $\boldsymbol{\alpha }$, respectively. Substituting \eqref{u_k} into \eqref{F_ij}, we have
\begin{equation}
	\begin{aligned}
		\frac{\partial \bar{\boldsymbol{u}}_k}{\partial \theta _T}=&\dot{\mathbf{a}}\left( \theta _T \right) \otimes \left[ \mathbf{b}_k\left( \Delta r \right) \odot \mathbf{c}\left( \theta _T \right) \right] 
		\\
		&+\mathbf{a}\left( \theta _T \right) \otimes \left[ \mathbf{b}_k\left( \Delta r \right) \odot \dot{\mathbf{c}}\left( \theta _T \right) \right]	, 
	\end{aligned}
\end{equation}
and
\begin{equation}
	\begin{aligned}
		\frac{\partial \bar{\boldsymbol{u}}_k}{\partial R_T}=\mathbf{a}\left( \theta _T \right) \otimes \left[ \dot{\mathbf{b}}_k\left( \Delta r \right) \odot \mathbf{c}\left( \theta _T \right) \right] 	,
	\end{aligned}
\end{equation}
where 
\begin{equation}
	\begin{aligned}
		\dot{\mathbf{a}}\left( \theta _T \right) =\frac{\partial \mathbf{a}\left( \theta _T \right)}{\partial \theta _T}=j2\pi \frac{d_3}{\lambda _0}\cos \theta _T\mathbf{E}_{\mathbf{a}}\mathbf{a}\left( \theta _T \right) 	,
	\end{aligned}
\end{equation}
\begin{equation}
	\begin{aligned}
		\dot{\mathbf{c}}\left( \theta _T \right) =\frac{\partial \mathbf{c}\left( \theta _T \right)}{\partial \theta _T}=j2\pi \frac{d_1}{\lambda _0}\cos \theta _T\mathbf{E}_{\mathbf{c}}\mathbf{c}\left( \theta _T \right) 	,
	\end{aligned}
\end{equation}
and
\begin{equation}
	\begin{aligned}
		\dot{\mathbf{b}}_k\left( \Delta r \right) =\frac{\partial \mathbf{b}_k\left( \Delta r \right)}{\partial R_T}=-j\frac{4\pi}{c}\mathbf{E}_{\mathbf{f}_k}\mathbf{b}_k\left( \Delta r \right) 	,
	\end{aligned}
\end{equation}
with $\mathbf{E}_{\mathbf{a}}=\mathrm{diag}\left\{ 0,\cdots ,M-1 \right\}$, $\mathbf{E}_{\mathbf{c}}=\mathrm{diag}\left\{ 0,\cdots ,N-1 \right\}$ and $\mathbf{E}_{\mathbf{f}_k}=\mathrm{diag} \left\{ \Delta f_{k,1}^{u,v},\cdots ,\Delta f_{k,N}^{u,v} \right\} $. 

For mathematical convenience, let $\boldsymbol{z}_{k}^{R_T}=\mathbf{Q}^{-\frac{1}{2}}\frac{\partial \bar{\boldsymbol{u}}_k}{\partial R_T}$, $\boldsymbol{z}_{k}^{\theta _T}=\mathbf{Q}^{-\frac{1}{2}}\frac{\partial \bar{\boldsymbol{u}}_k}{\partial \theta _T}$ and $\boldsymbol{z}_k=\mathbf{Q}^{-\frac{1}{2}}\bar{\boldsymbol{u}}_k$, the FIM for $k$ snapshots can be expressed as \eqref{F_Mat}, shown at the top of next page.  
\begin{figure*}[hbpt]
	\vspace{-0.5cm}
	{\small \begin{equation}\label{F_Mat}
			\begin{aligned}
				\mathbf{F}=\frac{2P_{\mathrm{S}}}{N}\mathrm{Re}\left\{ \left[ \begin{matrix}
					\sum\limits_{k=1}^K{\left\{ \left| \xi \right|^2\left\| \boldsymbol{z}_{k}^{R_T} \right\| ^2 \right\}}&		\sum\limits_{k=1}^K{\left\{ \left| \xi \right|^2\left( \boldsymbol{z}_{k}^{R_T} \right) ^H\boldsymbol{z}_{k}^{\theta _T} \right\}}&		\sum\limits_{k=1}^K{\left\{ \xi ^{\dagger}\left( \boldsymbol{z}_{k}^{R_T} \right) ^H\boldsymbol{z}_k \right\}}&		\sum\limits_{k=1}^K{\left\{ j\xi ^{\dagger}\left( \boldsymbol{z}_{k}^{R_T} \right) ^H\boldsymbol{z}_k \right\}}\\
					\sum\limits_{k=1}^K{\left\{ \left| \xi \right|^2\left( \boldsymbol{z}_{k}^{\theta _T} \right) ^H\boldsymbol{z}_{k}^{R_T} \right\}}&		\sum\limits_{k=1}^K{\left\{ \left| \xi \right|^2\left\| \boldsymbol{z}_{k}^{\theta _T} \right\| ^2 \right\}}&		\sum\limits_{k=1}^K{\left\{ \xi ^{\dagger}\left( \boldsymbol{z}_{k}^{\theta _T} \right) ^H\boldsymbol{z}_k \right\}}&		\sum\limits_{k=1}^K{\left\{ j\xi ^{\dagger}\left( \boldsymbol{z}_{k}^{\theta _T} \right) ^H\boldsymbol{z}_k \right\}}\\
					\sum\limits_{k=1}^K{\left\{ \xi (\boldsymbol{z}_k) ^H\boldsymbol{z}_{k}^{R_T} \right\}}&		\sum\limits_{k=1}^K{\left\{ \xi \left( \boldsymbol{z}_k \right) ^H\boldsymbol{z}_{k}^{\theta _T} \right\}}&		\sum\limits_{k=1}^K{\left\{ \left\| \boldsymbol{z}_k \right\| ^2 \right\}}&		0\\
					\sum\limits_{k=1}^K{\left\{ -j\xi \left( \boldsymbol{z}_k \right) ^H\boldsymbol{z}_{k}^{R_T} \right\}}&		\sum\limits_{k=1}^K{\left\{ -j\xi \left( \boldsymbol{z}_k \right) ^H\boldsymbol{z}_{k}^{\theta _T} \right\}}&		0&		\sum\limits_{k=1}^K{\left\{ \left\| \boldsymbol{z}_k \right\| ^2 \right\}}\\
				\end{matrix} \right] \right\} 	.
			\end{aligned}
	\end{equation}}
\end{figure*}
Define auxiliary variable $\mathbf{F}_{11}=\left[ \begin{matrix}
	F_{1,1}&		F_{1,2}\\
	F_{2,1}&		F_{2,2}\\
\end{matrix} \right] $, $\mathbf{F}_{12}=\left[ \begin{matrix}
	F_{1,3}&		F_{1,4}\\
	F_{2,3}&		F_{2,4}\\
\end{matrix} \right]$, $\mathbf{F}_{21}=\left[ \begin{matrix}
	F_{3,1}&		F_{3,2}\\
	F_{4,1}&		F_{4,2}\\
\end{matrix} \right]$, and $\mathbf{F}_{22}=\left[ \begin{matrix}
	F_{3,3}&		F_{3,4}\\
	F_{4,3}&		F_{4,4}\\
\end{matrix} \right]$, where $F_{i,j}$ represents the $(i,j)$th element of $\mathbf{F}$.
Using the block matrix inverse formula \cite{zhang2017matrix}, the inverse of FIM can be obtained as 
\begin{equation}
	\begin{aligned}
		\mathbf{\Lambda }=\mathbf{F}^{-1}=\frac{N}{2P_{\mathrm{S}}}\left[ \begin{matrix}
			\mathbf{D}^{-1}&		\times\\
			\times&		\times\\
		\end{matrix} \right]  	,
	\end{aligned}
\end{equation}
with
\begin{equation}
	\begin{aligned}
		\mathbf{D}=\mathbf{F}_{11}-\mathbf{F}_{12}\mathbf{F}_{22}^{-1}\mathbf{F}_{21}=\left[ \begin{matrix}
			D_{11}&		D_{12}\\
			D_{21}&		D_{22}\\
		\end{matrix} \right] 	
	\end{aligned}.
\end{equation}

Therefore, the CRBs for range and angle can be expressed as
\begin{equation}
	\begin{aligned}
		\mathrm{CRB}_{R_T}=\frac{N}{2P_{\mathrm{S}}}\frac{1}{\det \left( \mathbf{D} \right)}D_{22}	,
	\end{aligned}
\end{equation}
and
\begin{equation}
	\begin{aligned}
		\mathrm{CRB}_{\theta _T}=\frac{N}{2P_{\mathrm{S}}}\frac{1}{\det \left( \mathbf{D} \right)}D_{11}	,
	\end{aligned}
\end{equation}
respectively, where $\det \left( \cdot \right) $ denotes the determinant operator.

\section{Simulation Results}
\label{sc5}
In this section, we perform Monte Carlo simulations to evaluate the BER and the  root mean square error (RMSE) of the proposed system. Unless specified, the simulation parameters are set to $P_S=1$, $f_c=10\mathrm{GHz}$, $(R_C,\theta _C)=(300m,60\degree)$, $M=N$, $\Delta f=2\mathrm{MHz}$, $\sigma ^2=1$, $d_1=d_3=c/f_c$. $\theta _T$ and $R_T$ are randomly generated. The  SNR of the communication receiver and sensing receiver are defined as $\mathrm{SNR} =\frac{P\mathrm{s}}{N_0}$ \cite{basit2021fda,ji2019physical,jian2023mimo} and $\mathrm{SNR} =\frac{P\mathrm{s}}{NN_1}$ \cite{gui2017coherent,xu2015joint}, respectively. Note that ‘Ana’ and ‘Sim’ stand for the BER theoretical upper bound and the Monte Carlo simulation results in the following figures, respectively.

\subsection{Communication sub-function simulation}

\begin{figure*}[htbp]
	\centering
	\subfigure[]	{\includegraphics[width=0.32\textwidth]{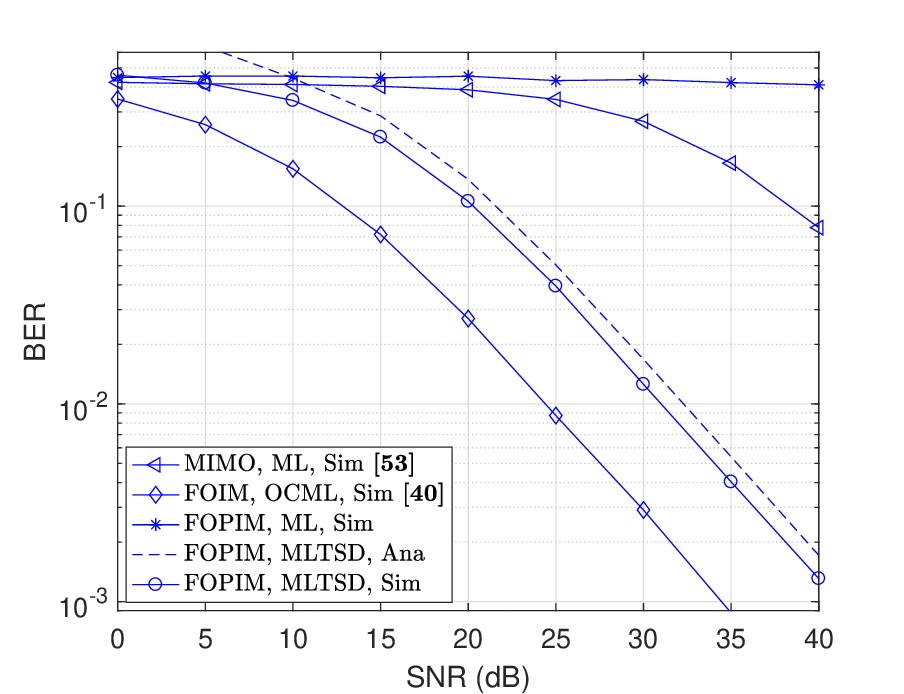}}\label{BER_Comp_1L}
	\subfigure[]	
	{\includegraphics[width=0.32\textwidth]{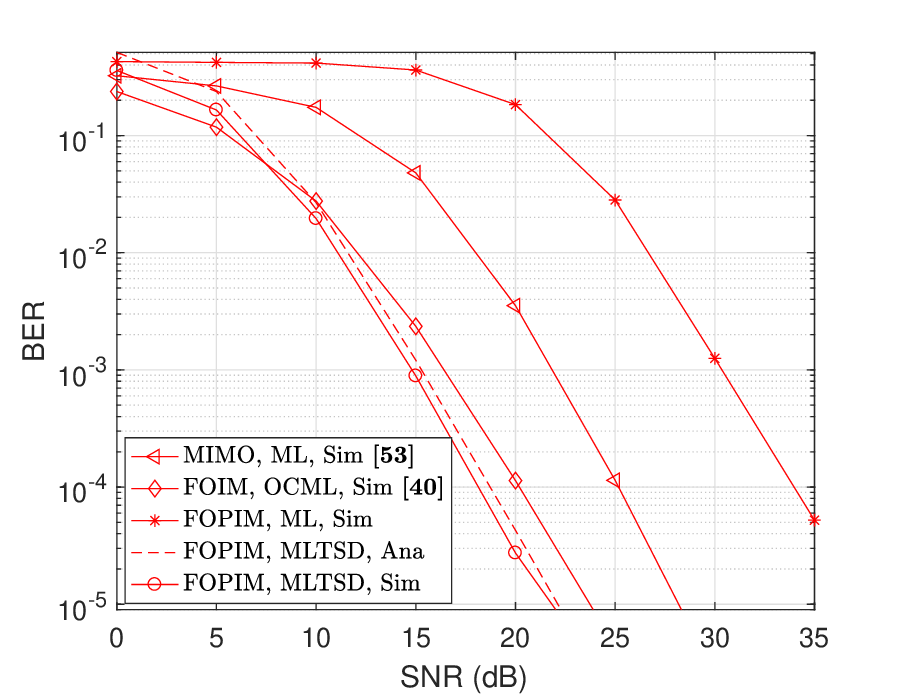}}
	\subfigure[]
	{\includegraphics[width=0.32\textwidth]{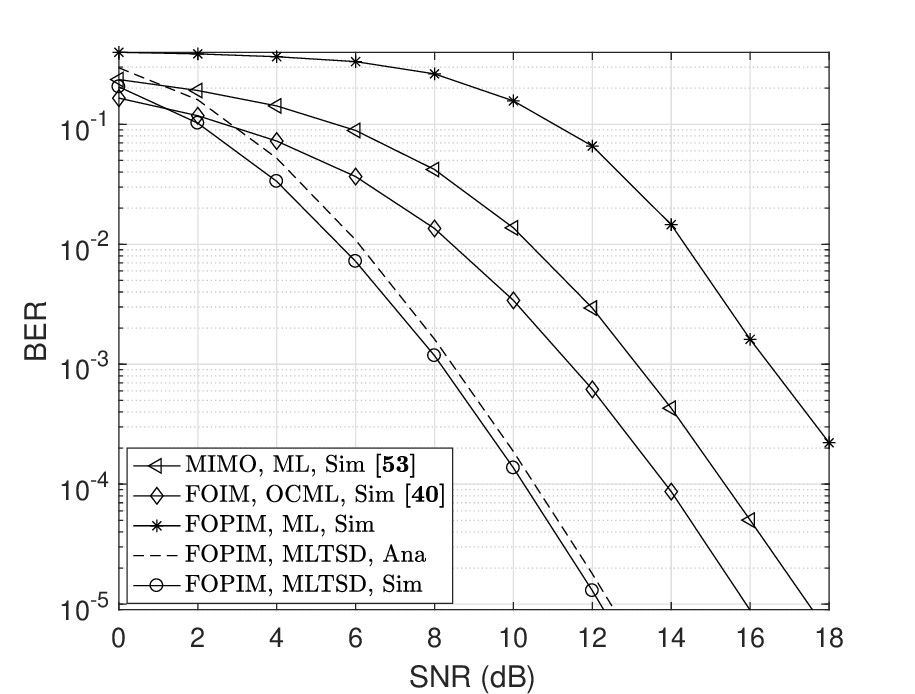}}
	\caption{BER performance comparison with different number of receive antennas, where $J=4$, $N=6$, $P=7$. (a) $L=1$. (b) $L=3$. (c) $L=6$.}
	\label{BER_Comp_136L}
\end{figure*}
In this subsection, the communication performance (BER, bits per pulse) of the proposed system is investigated. 

In Fig. \ref{BER_Comp_136L}, the BER performance of the proposed FOPIM, FOIM \cite{jian2023mimo} and traditional MIMO comunication schemes \cite{biglieri2007mimo} are compared for different numbers of receive antennas. The MIMO scheme shares the same number of antennas and modulation orders as the FOPIM method.
Note that the derivation of decoding FOPIM by the ML method is provided in Appendix \ref{ApxA}. For the comparison fairness, the emitted bits of the FOIM and FOPIM schemes are set to be close. That is, $J=8$ is employed in the FOIM scheme \cite{jian2023mimo}, with all other parameters identical to the FOPIM scheme.
The simulation results match well with the analysis results, which verify the theoretical upper bound. 
Moreover, the BER performance of the MIMO scheme outperforms the FOPIM scheme with ML decoding, but is inferior to the FOPIM scheme using MLTSD.
This is because with the ML method, the FOPIM approach necessitates $J^N \times 2^{\lfloor \log _2N! \rfloor}$ searches, causing a smaller decision domain than FOIM approach. Recall \eqref{FOPIM_ML}, the MLTSD method individually estimates the constellation symbol of each antenna through ML criteria, thereby diminishing the decoding errors.

Another interesting observation in Fig. \ref{BER_Comp_136L} is that the FOIM method exhibits superior BER compared to the FOPIM method when $L=1$, while FOPIM outperforms FOIM with a larger $L$. This can be explained as follows. In \eqref{Pr_hh}, when $L=1$, $P_{\mathcal{P}}$ becomes high due to the absence of receive diversity gain. This leads to a higher error probability in the index and constellation estimates compared to the FOIM scheme \cite{jian2023mimo}. Comparing \eqref{P_P} with (51) in \cite{jian2023mimo} reveals that the $P_I$ and $P_{\mathcal{P}}$ sharply decrease with the increasing $L$, and consequently, FPOIM has a smaller antenna index estimation error probability.
Moreover, \eqref{P_e} demonstrates that the frequency offset index estimation error probability increases slightly with higher $J$. In the simulation, FOIM uses $J=8$, while FOPIM uses $J=8$. This leads to a higher probability of constellation symbol error and index estimation error for FOIM.

\begin{figure}[htbp]
	\vspace{-0.5cm}
	\centering
	\subfigure	{\includegraphics[width=0.35\textwidth]{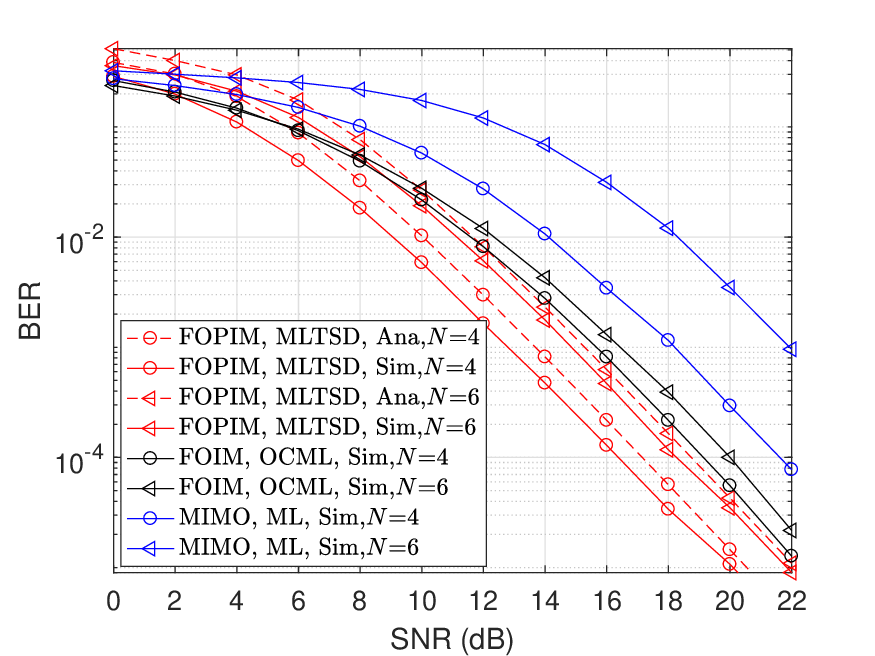}}
	\caption{BER performance comparison with different number of transmit antennas, where $J=4$, $L=3$, $P=7$.}
	\label{BER_VS_diff_N}
\end{figure}
Fig. \ref{BER_VS_diff_N} shows the variation of BER with $N$ for different communication schemes. To achieve a similar information bits carryover as the FOPIM scheme, the FOIM scheme was set to $J = 8$. We observe that the proposed FOPIM scheme showcases over 5 dB superior BER performance compared to the MIMO system at a BER of $4.7\times 10^{-4}$. 
The BER of the FOPIM method is superior to that of the FOIM scheme when $\mathrm{SNR}>8$dB. The reason is that to achieve a similar amount of information, the FOIM approach requires a higher modulation order, which suffers a higher constellation error probability. In addition, as $N$ rises, the judge domain for decoding shrinks, resulting in a deterioration of the system BER.

\begin{figure}[htbp]
	\centering
	\subfigure	{\includegraphics[width=0.35\textwidth]{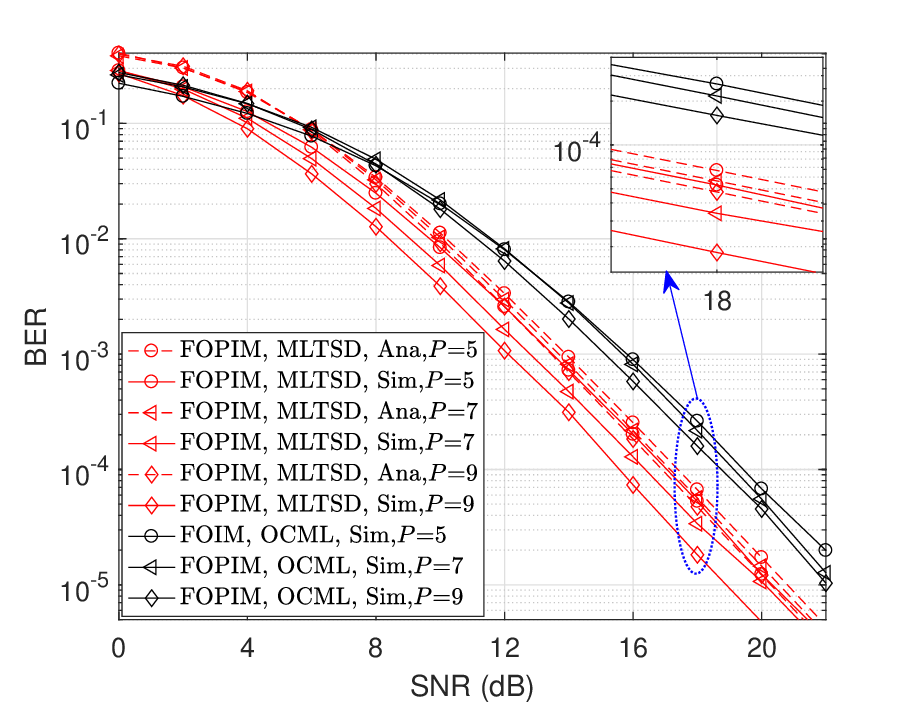}}
	\caption{BER performance comparison with different frequency offset pool size, where $N=4$, $L=3$, $J=4$.}
	\label{BER_VS_diff_P}
\end{figure}
In Fig. \ref{BER_VS_diff_P}, we investigate the BER of the FOPIM and FOIM methods with the different frequency offset pool sizes. Note that $J=8$ is applied to the FOIM scheme for comparison fairness. Observing Fig. \ref{BER_VS_diff_P} and Fig. \ref{BER_VS_diff_N} depicts that the gap between analyzed and simulated results slightly expands with the discrepancy between $N$ and $P$. This is because the number of decision domains increases as $P$ increases, intensifying the overlap among these domains in \eqref{P_r_orig} during PEP calculation. 
The proposed method exhibits a lower BER than FOIM across different $P$. The reason is that the FOPIM approach can carry more indexed bits for the same information content, reducing the error probability related to constellation symbols. Another noteworthy discovery is the slight decrease in the system's BER with the rise in $P$. This can be elucidated by revisiting \eqref{y_wave_lp}, where the noise power at each parallel filter output diminishes as $P$ increases. Consequently, this leads to a reduction in the error probability of \eqref{fre_est} and \eqref{FOPIM_ML}. 

\begin{figure}[htb]
	\centering
	\subfigure[]	{\includegraphics[width=0.35\textwidth]{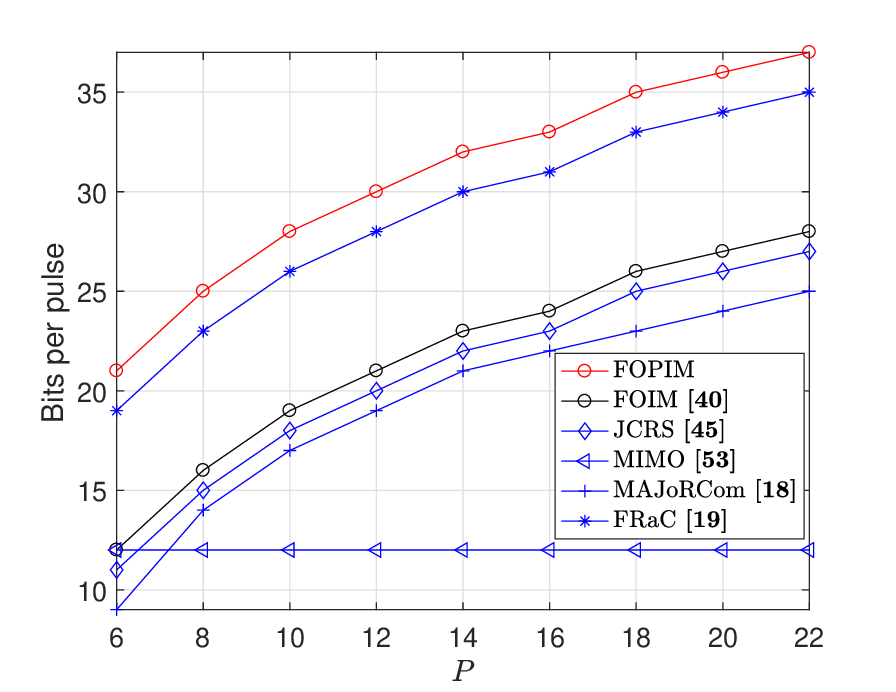}}\label{rate_comp_diff_P}
	\subfigure[]	
	{\includegraphics[width=0.35\textwidth]{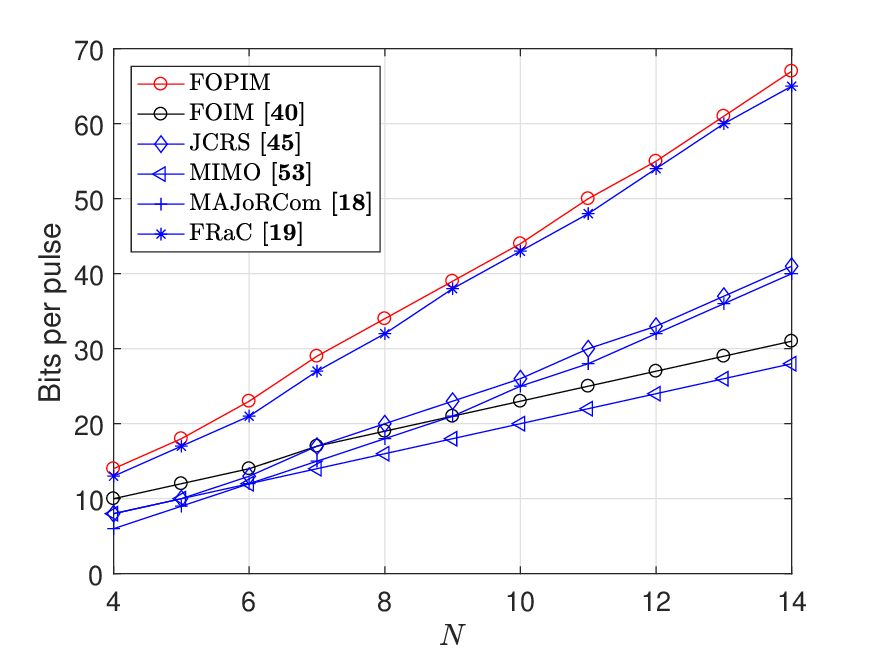}}\label{rate_comp_diff_N}
	\caption{Bits per pulse comparison among different schemes, where $J=4$, $N=6$. (a) Bits per pulse versus frequency pool size. (b) Bits per pulse versus the number of antennas, where $P=N+1$.}
	\label{rate_comp_diff_PN}
\end{figure}
Fig. \ref{rate_comp_diff_PN} compares the bits per pulse across different schemes: proposed FOPIM, FOIM \cite{jian2023mimo}, MIMO \cite{biglieri2007mimo}, JCRS \cite{li2023joint}, MAJoRCom \cite{huang2020majorcom}, and FRaC \cite{ma2021frac}. Note that JCRS's waveform set size equals $P$, MAJoRCom employs a separate frequency for each antenna, and FRaC actives $N-2$ antennas. Inspecting Fig. \ref{rate_comp_diff_PN} reveals that the FRaC scheme carries more bits than the JCRS, FOIM, MAJoRCom, and MIMO schemes, which is attributed to its multimode indexed modulation involving joint array, frequency offset, and constellation. Interestingly, bits carried by proposed FOPIM approach set an upper limit for FRaC's, notably in Fig. \ref{rate_comp_diff_PN}(a) with a 2-bit margin. This phenomenon can be elaborated as follows: the FRaC carries $k_{\mathrm{FRaC}}=\lfloor \log _2C_{N}^{N_1} \rfloor +\lfloor \log _2C_{P}^{N_1} \rfloor +\lfloor \log _2N_1! \rfloor +N_1\log _2J$ bits per pulse \cite{ma2021frac}, where $N_1$ denotes active antenna number. The bits carried by the FRaC method governed by the $\lfloor \log _2N_1! \rfloor +N_1\log _2J$ term, which is monotonically increasing with $N_1$. 
$k_{\mathrm{FRaC}}$ is maximized when $N_1$ takes the maximum value, i.e., $N_1=N$, and then the bits carried by the FRaC scheme is equal to the FOPIM scheme.

\subsection{Sensing sub-function simulation}
In this subsection, we study the sensing performance of the proposed FDA-MIMO ISAC system with the FOPIM scheme (denoted as "FOPIM" in simulations). The MIMO system, commonly employed in existing ISAC systems \cite{liu2020joint,chen2022generalized,hassanien2015dual,liu2022transmit}, is simulated for comparison. Note that the MIMO scheme's bandwidth is configured to match the total bandwidth of the FOPIM scheme, denoted as $P \Delta f$. The simulation results obtained by 400 independent Monte Carlo trails.

\begin{figure}[htbp]
	\centering
	\subfigure[]	{\includegraphics[width=0.35\textwidth]{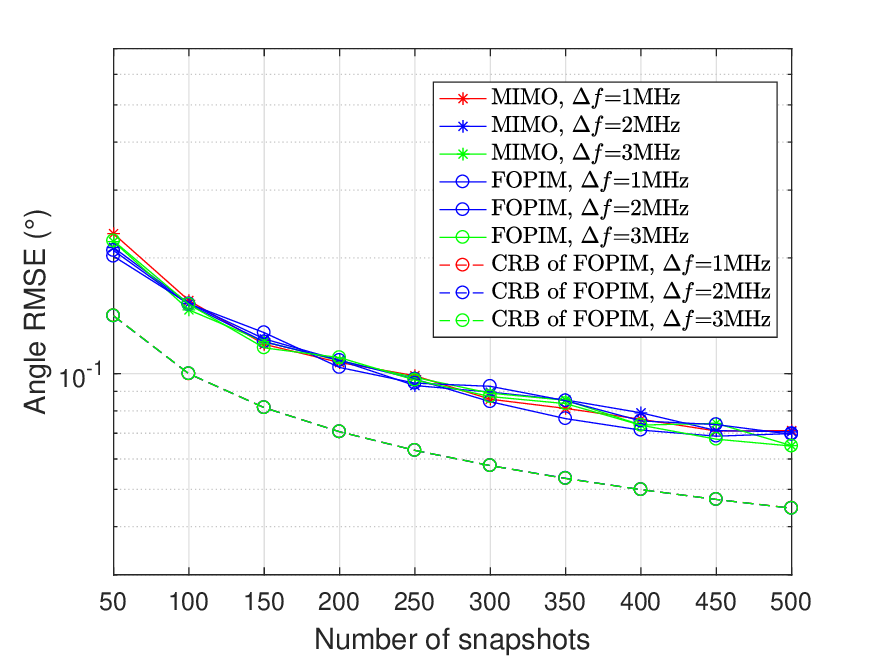}}\label{RMSE_VS_Snaps_dif_deltaF_angle}
	\subfigure[]	
	{\includegraphics[width=0.35\textwidth]{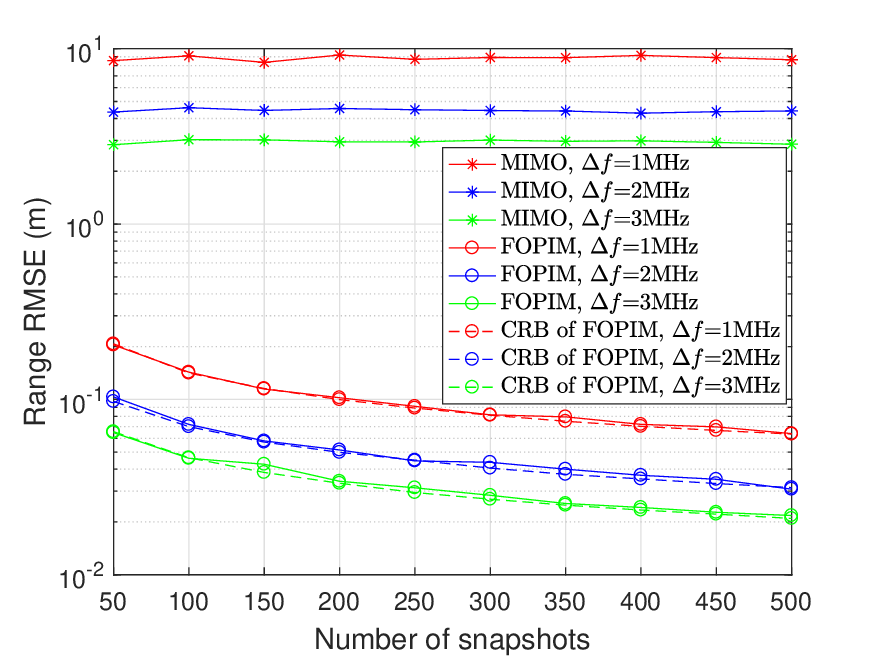}}\label{RMSE_VS_Snaps_dif_deltaF_range}
	\caption{RMSE versus snapshots with different frequency offsets, where $N=6$, $P=7$, $\mathrm{SNR}=0\mathrm{dB}$. (a) Angle estimation. (b) Range estimation.}
	\label{RMSE_VS_Snaps_dif_deltaF_angle_range}
\end{figure}

\begin{figure}[htbp]
	\centering
	\subfigure[]	{\includegraphics[width=0.35\textwidth]{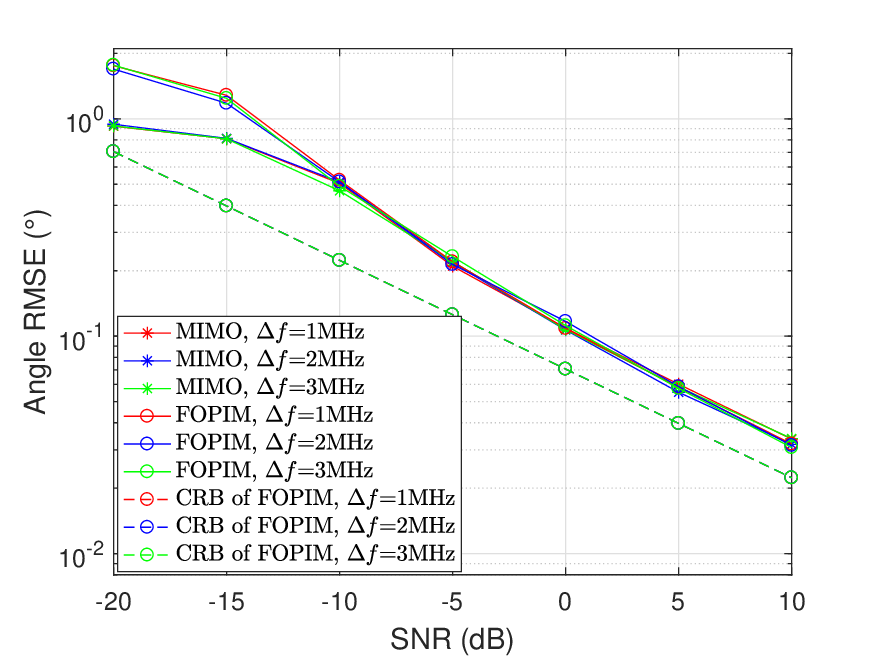}}\label{RMSE_VS_SNR_dif_deltaF_angle}
	\subfigure[]	
	{\includegraphics[width=0.35\textwidth]{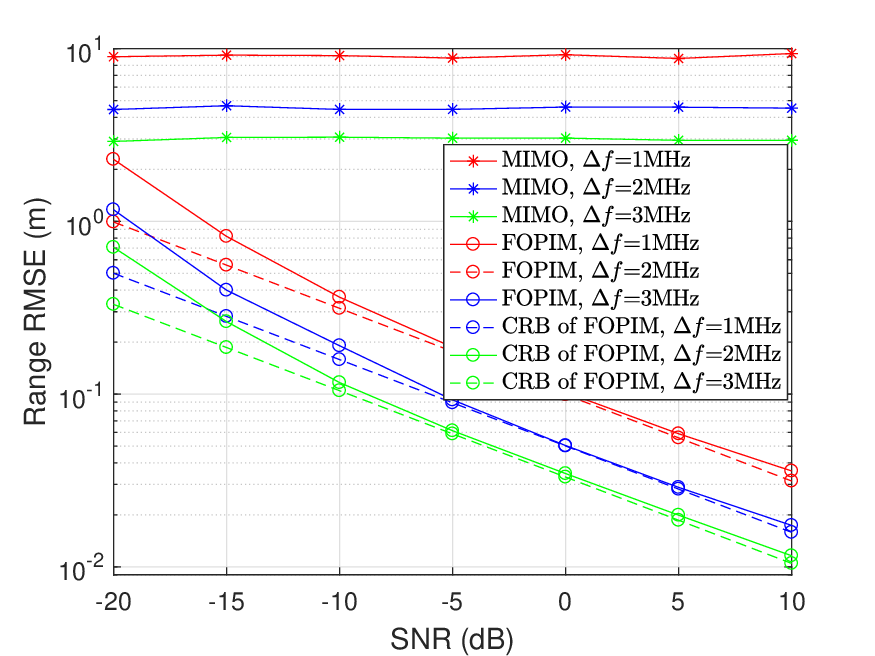}}\label{RMSE_VS_SNR_dif_deltaF_range}
	\caption{RMSE versus SNR with different frequency offsets, where $N=6$, $P=7$, $K=200$. (a) Angle estimation. (b) Range estimation.}
	\label{RMSE_VS_SNR_dif_deltaF_angle_range}
\end{figure}

Figs. \ref{RMSE_VS_Snaps_dif_deltaF_angle_range} and \ref{RMSE_VS_SNR_dif_deltaF_angle_range} show RMSEs and root CRBs for angle and range estimations in the proposed FOPIM scheme with varied frequency offsets. Note that the TSMLE method is also employed in the MIMO scheme. It is observed from Figs \ref{RMSE_VS_Snaps_dif_deltaF_angle_range} and \ref{RMSE_VS_SNR_dif_deltaF_angle_range} that the accuracies of angle and distance estimation is improved with the increasing snapshot count ($K$) and SNR. Moreover, the range estimation accuracies of both the proposed system and the MIMO system are reinforced as the frequency offset increases, which is attributed to the more significant frequency offset increasing the system bandwidth. Moreover, Fig. \ref{RMSE_VS_Snaps_dif_deltaF_angle_range}(b) reveals that increasing the number of snapshots from 50 to 500 results in a modest enhancement of only $0.17 \degree$ and 0.14 meters in angle and distance estimation accuracy, respectively. Referring to \ref{RMSE_VS_SNR_dif_deltaF_angle_range}(b), it is suggested that augmenting SNR is more effective for enhancing estimation accuracy.

Another observation is that the MIMO system and the proposed system have similar angle RMSE. In contrast, the proposed system has a much higher estimation accuracy in the range dimension than the MIMO system. This phenomenon is explained as follows: The MIMO and FDA-MIMO systems possess comparable apertures. However, the MIMO steering vector lacks distance information and can only randomly estimate the position of the target in the range bin after pulse compression. In contrast, the FDA-MIMO steering vector incorporates range information, enabling it to offer super-resolution capability in the range dimension \cite{gui2017coherent,xu2015joint}.

\begin{figure}[htbp]
	\centering
	\subfigure[]	{\includegraphics[width=0.35\textwidth]{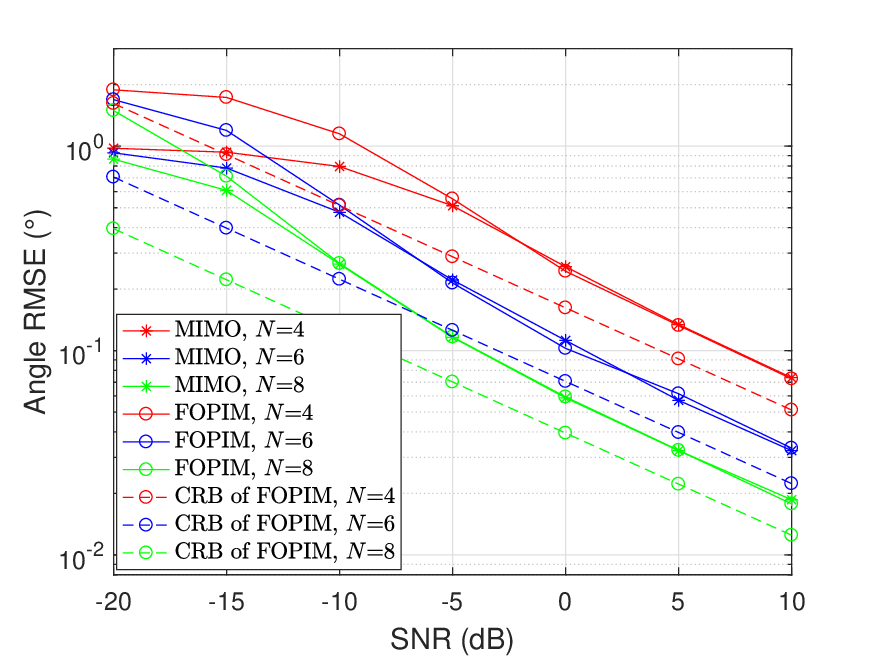}}\label{RMSE_VS_SNR_dif_N_angle}
	\subfigure[]	
	{\includegraphics[width=0.35\textwidth]{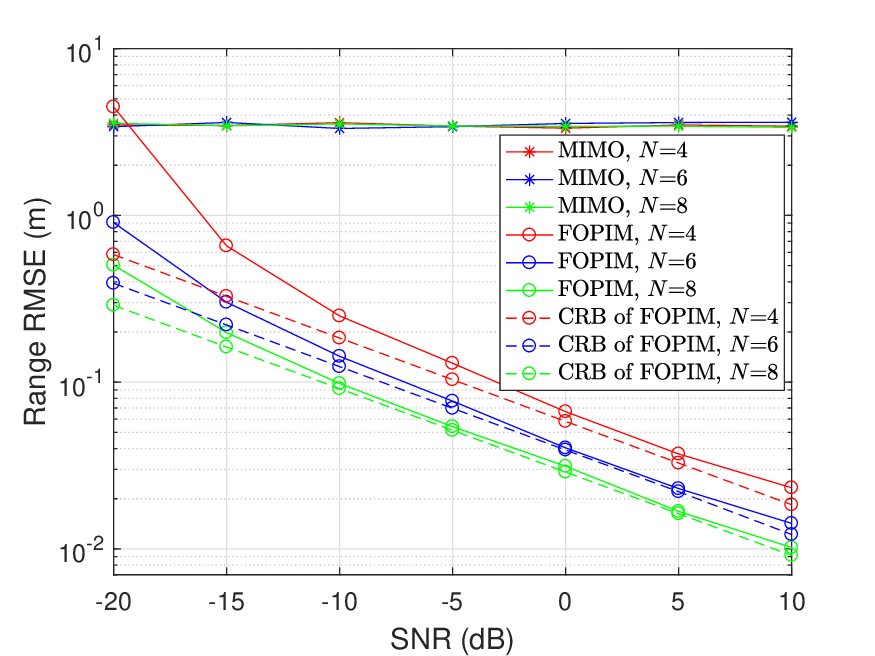}}\label{RMSE_VS_SNR_dif_N_range}
	\caption{RMSE versus SNR with different number of transmit antennas, where $\Delta f=\mathrm{2MHz}$, $P=9$, $K=200$. (a) Angle estimation. (b) Range estimation.}
	\label{RMSE_VS_SNR_dif_N_angle_range}
\end{figure}

Fig. \ref{RMSE_VS_SNR_dif_N_angle_range} illustrates RMSEs and root CRBs for the proposed FOPIM scheme and MIMO scheme with different numbers of transmit antennas. The angle estimation RMSE of MIMO and FOPIM are improved with increasing $N$. Moreover, as $N$ rises, the range RMSE of the MIMO approach stands at a high level (around 3.45m), whereas the proposed system exhibits a diminishing range RMSE, reaching below 0.001m at 10dB. This is because the virtual aperture of the FDA-MIMO system serves for both angle and range estimation, while the MIMO system uses it solely for angle estimation. Increasing $N$ expands the system's virtual aperture, leading to higher accuracy in the angle and distance estimation of the proposed system, whereas the unchanged range precision and higher angle precision of the MIMO system.

\begin{figure}[htbp]
	\centering
	\subfigure[]	{\includegraphics[width=0.35\textwidth]{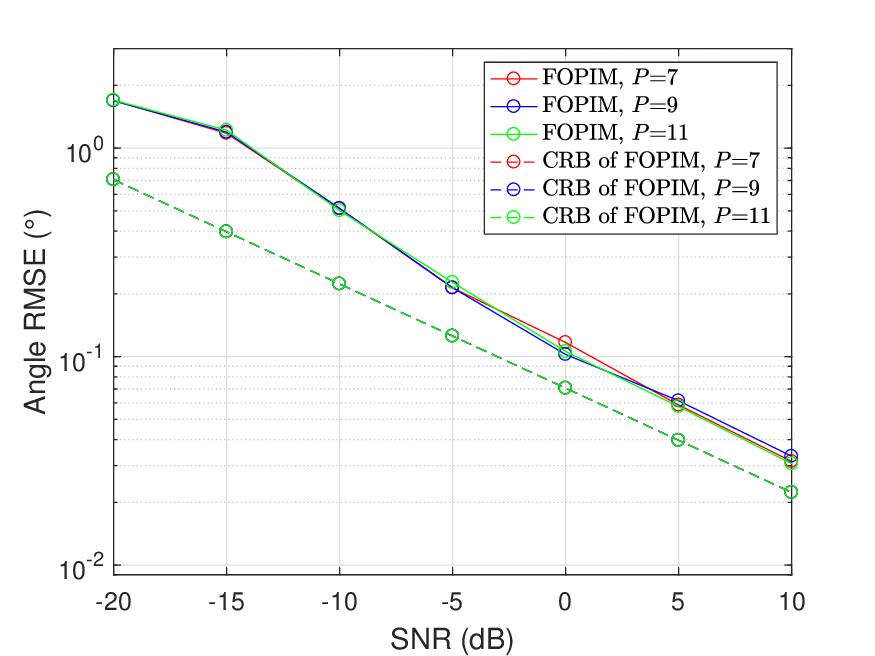}}\label{RMSE_VS_SNR_dif_P_angle}
	\subfigure[]	
	{\includegraphics[width=0.35\textwidth]{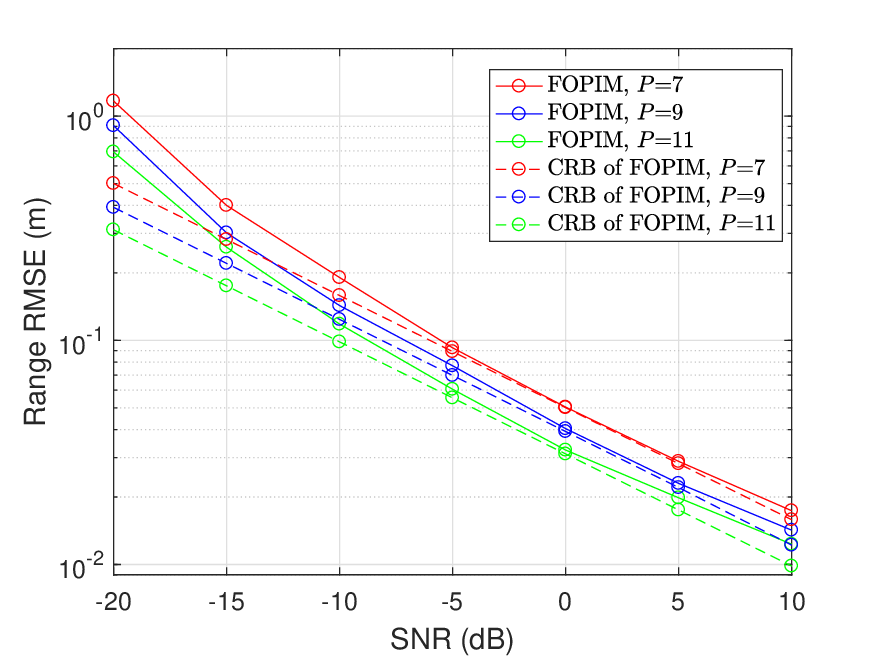}}\label{RMSE_VS_SNR_dif_P_range}
	\caption{RMSE versus SNR with different frequency offset pool sizes, where $\Delta f=\mathrm{2MHz}$, $N=6$, $K=200$. (a) Angle estimation. (b) Range estimation.}
	\label{RMSE_VS_SNR_dif_P_angle_range}
\end{figure}

Fig. \ref{RMSE_VS_SNR_dif_P_angle_range} examines the impact of various frequency offset pool sizes on the RMSE and root CRB of the proposed system. One can find that the angle RMSE and angle root CRB remain unchanged with varying $P$, since the array aperture remains constant. Interestingly, the range RMSE and range root CRB decreases with increasing $P$. This phenomenon can be explained by the fact that the total bandwidth of the system increases as $P$ grows, leading to an enforcement in the range estimation accuracy.

\section{Conclusion}
\label{sc6}
In this paper, the FOPIM scheme was proposed for the FDA-MIMO-based ISAC system, which enhanced communication capability by selecting and permutating frequency offsets to deliver extra information bits. Considering the communication user suffers from high decoding complexity and BER, an MLTSD method was proposed to tackle this issue. The MLTSD method first detected selected frequency offsets by searching maximum outputs, then utilized ML criterion to estimate the indices of frequency offsets and corresponding QAM symbols. We employed the TSMLE to estimate the target angle and range stepwise for the system sensing sub-function. Moreover, closed-form expressions were derived for the tight upper bound on the BER of the communication sub-funtion and for the CRB of the sensing sub-funtion. Simulation results corroborated the theoretical analysis, showing that the proposed system achieved lower BER than the standalone MIMO and FOIM schemes, as well as the superior range resolution than the independent sensing module.

\appendices
\section{}
\label{ApxA}
In a ML manner similar to \cite{ji2019physical,jian2023physical,basit2021fda,nusenu2020space}, the outputs of all the parallel filters for each antenna are summed. Recalling \eqref{y_p_wave}, the output summation signal of the $l$th antenna can be represented as
\begin{equation}
	\begin{aligned}
		\tilde{y}_l=\sum_{p=1}^P{\tilde{y}_{l,p}}=\sqrt{\frac{P_S}{N}}\mathring{\mathbf{h}}_{n}^{T}\mathbf{Ax}+n_l	,
	\end{aligned}
\end{equation}
where $n_l\sim \mathcal{C} \mathcal{N} \left( 0,N_0 \right) $ denotes the total receive noise power of the $l$th antenna,
\begin{equation}
	\begin{aligned}
		\mathbf{A}=\mathrm{diag}\left\{ e^{-j2\pi (f_c+\Delta f_{1}^{u,v})\tau _1},e^{-j2\pi (f_c+\Delta f_{2}^{u,v})\tau _2}, \right. 
		\\
		\left. \cdots ,e^{-j2\pi (f_c+\Delta f_{N}^{u,v})\tau _N} \right\}	,
	\end{aligned}
\end{equation}
and $\mathbf{x}=\left[ x_1,x_2,\cdots ,x_N \right] ^T$ stands for the transmitted constellation symbol vector.

Then, the summation outputs of $L$ receive antennas can be written as
\begin{equation}
	\begin{aligned}
		\tilde{\mathbf{y}}&=\left[ \tilde{y}_1,\cdots,\tilde{y}_l,\cdots ,\tilde{y}_L \right] 
		\\
		&=\sqrt{\frac{P_S}{N}}\mathbf{HGx}+\mathbf{n}	,
	\end{aligned}
\end{equation}
where $\mathbf{G}_{u,v}=\mathbf{SA}_{u,v}=[ \mathbf{g}_1,\cdots,\mathbf{g}_n,\cdots ,\mathbf{g}_N ]$. Note that only the $(n+N\Delta f_{n}^{u,v}/\Delta f)$th element in $\mathbf{g}_n=[0,\cdots ,e^{-j2\pi \left( f_c+\Delta f_{n}^{u,v} \right) \tau _n},\cdots ,0]^T$ is non-zero.

Further, the ML detector can be expressed as \cite{ji2019physical,basit2021fda}
\begin{equation}
	\begin{aligned}
		[\hat{\mathbf{G}},\hat{\mathbf{x}}]=\mathop {arg\min} \limits_{\mathbf{G},\mathbf{x}}\left\| \tilde{\mathbf{y}}-\sqrt{\frac{P_S}{N}}\mathbf{HGx} \right\| ^2		,
	\end{aligned}
\end{equation}
where $\hat{\mathbf{G}}$ and $\hat{\mathbf{x}}$ denote the estimations of $\mathbf{G}$ and $\mathbf{x}$, respectively. Relying on the receiver characteristics, the ML method can estimate emitted frequency offsets using the procedure of \eqref{y_p_wave} and \eqref{fre_est}. Consequently, the ML method solely requires traversing the estimated frequency offsets permutation in \eqref{fre_est} and constellation symbols permutation.

\ifCLASSOPTIONcaptionsoff
\newpage
\fi

\bibliographystyle{IEEEtran}

\end{document}